\documentclass[journal]{IEEEtran}
\usepackage[version=4]{mhchem}
\usepackage{amssymb}     
\usepackage{amsfonts}
\usepackage{amsmath}
\usepackage[dvips]{graphicx}
\usepackage{color}       
\usepackage{textcomp}
\usepackage{multirow}
\usepackage{txfonts}
\usepackage{hyperref}
\usepackage{textcomp}
\usepackage{mathtools}
\usepackage{lipsum}
\usepackage{lineno}
\usepackage{cite}
\UseRawInputEncoding
\usepackage{stfloats}
\usepackage{upgreek}
\usepackage{mhchem}
\def\degree{${}^{\circ}$}

\begin{document}

\title{Calibration of cascaded phase shifters using pairwise scan method in silicon photonics integrated chip}
\author{{Yanxiang Jia, Xuyang Wang$^*$, Yizhuo Hou, Yu Zhang, Yuqi Shi, Qiang Zhang, Jun Zou, and Yongmin Li$^*$}
\thanks{This work was supported in part by the Provincial Natural Science Foundation of Shanxi, China (Grant No. 202103021224010), in part by Shanxi Provincial Foundation for Returned Scholars, China (Grant No. 2022-016), in part by National Natural Science Foundation of China (Grant Nos. 62175138, 62205188, 11904219), in part by the ``1331 Project" for Key Subject Construction of Shanxi Province, China, in part by the Innovation Program for Quantum Science and Technology (Grant No. 2021ZD0300703). \emph{(Corresponding author: Xu-Yang Wang; Yong-Min Li.)}}
\thanks{Xuyang Wang and Yongmin Li are with the State Key Laboratory of Quantum Optics and Quantum Optics Devices, Institute of Opto-Electronics, Shanxi University, Taiyuan 030006, China, also with the Collaborative Innovation Center of Extreme Optics, Shanxi University, Taiyuan 030006, China, and also with the Hefei National Laboratory, Hefei 230088, China (e-mail: wangxuyang@sxu.edu.cn; yongmin@sxu.edu.cn).}
\thanks{Yanxiang Jia, Yizhuo Hou, Yu Zhang, and Yuqi Shi are with the State Key Laboratory of Quantum Optics and Quantum Optics Devices, Institute of Opto-Electronics, Shanxi University, Taiyuan 030006, China (e-mail: jiayanxiang5806@163.com; sxhyz8828@126.com; 1143810783@qq.com; 1187896575@qq.com).}
\thanks{Qiang Zhang is with the State Key Laboratory of Quantum Optics and Quantum Optics Devices, Institute of Opto-Electronics, Shanxi University, Taiyuan 030006, China, also with the Collaborative Innovation Center of Extreme Optics, Shanxi University, Taiyuan 030006, China (e-mail: qzhang@sxu.edu.cn).}
\thanks{Jun Zou is with the ZJU-Hangzhou Global Scientific and Technological Innovation Center, Zhejiang University, Hangzhou 311215, China (e-mail: junzou\_optics@zju.edu.cn).}
\thanks{}}

\maketitle

\begin{abstract}
Cascaded phase shifters (CPSs) based on silicon photonics integrated chips play important roles in quantum information processing tasks. Owing to an increase in the scale of silicon photonics chips, the time required to calibrate various CPSs has increased. We propose a pairwise scan method for rapidly calibrating CPSs by introducing equivalent Mach Zehnder Interferometer structures and a reasonable constraint of the initial relative phase. The calibration can be nearly completed when the scanning process is finished, and only a little calculation is required. To achieve better performance, the key components, thermal optical phase shifter and $2 \times 2$ 50/50 multimode interference coupler, were simulated and optimized to prevent thermal crosstalk and ensure a good balance. A 6-CPSs structure in a packaged silicon photonics chip under different temperature was used to verify the rapid pairwise scan method, and a fidelity of 99.97{$\%$} was achieved.
\end{abstract}

\begin{IEEEkeywords}
Integrated optics, Quantum information, Cascaded phase shifters, Mach Zehnder Interferometer, Fidelity.
\end{IEEEkeywords}

\section{Introduction}
\IEEEPARstart{S}{ilicon} photonics integrated chips offer attractive advantages in quantum information processing tasks \cite{ref1,ref2,ref3,ref4,ref5,ref6,ref7}, such as large-scale integration \cite{ref8,ref9,ref10}, complementary metal-oxide semiconductor compatibility \cite{ref11,ref12}, telecommunication wavelengths operation \cite{ref13}, integrated photon sources \cite{ref14,ref15,ref16}, programmable quantum circuits \cite{ref17,ref18,ref19}, quantum key distribution \cite{sup10,sup8,sup11,sup12}, and integrated photon detectors \cite{ref20,ref21,ref22}. To ensure correct operation of the quantum circuits, the calibration of the components in the quantum circuits is necessary. The calibration of numerous phase shifters has become increasingly complex owing to an increase in the scale of quantum photonics chips. A representative example is the cascaded phase shifters (CPSs) in a fully programmable two-qubit quantum processor \cite{ref3,ref23}. There are a total of eight 5-CPSs in the quantum processor. Each 5-CPSs can be used to prepare initial single-qubit states and implement single-qubit operation. The structure of CPSs comprises a series of $2 \times 2$ 50/50 multimode interference (MMI) couplers and parallel waveguides with phase shifters between them. By varying the relative phase of the two parallel waveguides, the specified function can be accomplished. However, owing to fabrication errors, the initial relative phase is usually not zero, and the linear relationship between the heating power and shifted phase for each phase shifter is not identical. Thus, CPSs should be calibrated prior to use in a quantum information processing task.

\indent The traditional method for calibrating the CPSs structure requires a lot of calculation, which increases exponentially with the series number $N$ of CPSs \cite{ref3}. Later, a calibration method whose amount of calculation linearly increasing with the series number $N$ was proposed and demonstrated\cite{ref24,ref25}. In this paper, we propose a pairwise scan method for calibrating CPSs. The proposed method introduced an equivalent transform of Mach Zehnder Interferometer (MZI) structures that can simplify CPSs structures. Using this method, CPSs can be calibrated by scanning pairs of thermal optical phase shifters (TOPSs) stepwise. After scanning, the calibration is nearly completed, only a little calculation is required. To simplify the calibration, the pairwise scan method utilizes the constraint that the initial relative phase, $\Delta \theta $, satisfies $\left| {\Delta \theta } \right| < {\pi  \mathord{\left/{\vphantom {\pi  2}} \right.\kern-\nulldelimiterspace} 2}$, which is a reasonable assumption for mature fabrication technology. A method for calibrating CPSs without the constraint is also proposed and demonstrated. To reasonably design the TOPS, we simulated the temperature distribution around TOPS on the silicon-on-insulator (SOI) platform, a safe distance is designed to prevent the thermal crosstalk between TOPSs. Furthermore, the $2 \times 2$ 50/50 MMI structure was optimized to reduce calibration error. Finally, the average fidelity of 99.97$\%$ was achieved by using a 6-CPSs structure in a packaged silicon photonics chip.

\section{CPSs calibration methods}
During the calibration of CPSs, we discriminate the methods as odd or even pairwise scan methods according to the series $N$ of phase shifters.

\indent In the odd pairwise scan method, $N$ is expressed as
\begin{equation}
	N = 2n - 1,n \in {\mathbb{Z}^ + }. 
\end{equation}
\indent In the even pairwise scan method,  $N$ is expressed as
\begin{equation}
	N = 2n,n \in {\mathbb{Z}^ + }. 
\end{equation}
\indent The structure of 1-CPS is the simplest among the odd CPSs, and the structure of 2-CPSs is the simplest among the even CPSs. They are both fundamental structures. In the following part, we will discuss the calibrations of 1-CPS, 2-CPSs, even CPSs, and odd CPSs.

\subsection{Calibration of 1-CPS}
\begin{figure}[ph]
	\vspace{-0.3cm}
	\centering
	\includegraphics[width=0.67\columnwidth]{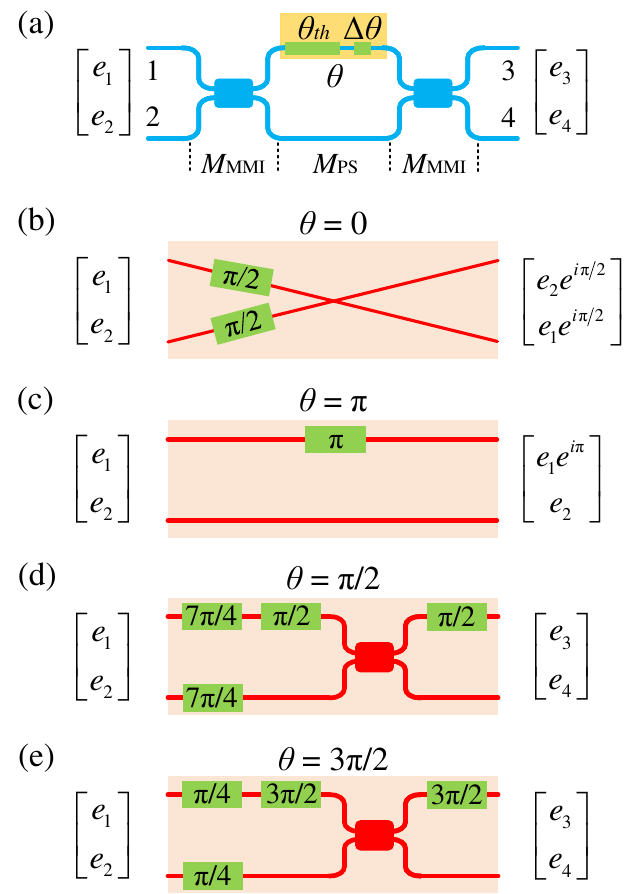}
	\caption{Structure of 1-CPS and its equivalent structures. (a) Structure of 1-CPS. (b) Equivalent cross-connection structure when $\theta  = 0$. (c) Equivalent direct-connection structure when $\theta  = \pi$. (d) Equivalent structure when $\theta  = {\pi  \mathord{\left/{\vphantom {\pi  2}} \right.\kern-\nulldelimiterspace} 2}$. (e) Equivalent structure when $\theta  = 3{\pi  \mathord{\left/ {\vphantom {\pi  2}} \right. \kern-\nulldelimiterspace} 2}$. } 
	\label{fig1}
	\vspace{-0.3cm}
\end{figure}

Fig. \ref{fig1}(a). shows the structure of 1-CPS, which is a typical MZI structure that has been widely used in silicon photonics \cite{sup1,sup2,sup3,sup4,sup5,sup6,sup7,ref17}. The initial relative phase of the two parallel waveguides is $\Delta \theta $, and the relative phase generated by TOPS is ${\theta _{th}}$, which has a linear relationship with the heating power $P$ that applied. The total relative phase $\theta $ is expressed as         
\begin{equation}
	\theta  = {\theta _{th}} + \Delta \theta  = kP + \Delta \theta . \label{Eq3}
\end{equation}
The Jones Matrix of 1-CPS is represented as 
\begin{equation}
	\begin{aligned}
		M_{\mathrm{MZI}} & =M_{\mathrm{MMI}} \cdot M_{\mathrm{ps}}(\theta) \cdot M_{\mathrm{MMI}} \\
		& =\frac{1}{2}\left[\begin{array}{cc}
			-1+e^{i \theta} & i\left(1+e^{i \theta}\right) \\
			i\left(1+e^{i \theta}\right) & 1-e^{i \theta}
		\end{array}\right],
	\end{aligned}
\end{equation}
where ${M_{{\text{MMI}}}}$ is the Jones Matrix of $2 \times 2$ MMI, and ${M_{{\text{ps}}}}\left( \theta  \right)$ is the Jones Matrix of phase shift $\theta $ in upper arm.

\indent The laser beams guided into ports 1 and 2 are denoted as ${e_1} = \left| {{e_1}} \right|{e^{{i\phi _1}}}$ and ${e_2} = \left| {{e_2}} \right|{e^{{i\phi _2}}}$, respectively. Their normalized intensities satisfy the following equation
\begin{equation}
{I_1} + {I_2} = {\left| {{e_1}} \right|^2} + {\left| {{e_2}} \right|^2} = 1.
\end{equation} 
The Jones vector of the input beams can be expressed as      
\begin{equation}
	{e_{in}} = \left[ {\begin{array}{*{20}{c}}
			{{e_1}} \\ 
			{{e_2}} 
	\end{array}} \right] = \left[ {\begin{array}{*{20}{c}}
			{\left| {{e_1}} \right|{e^{i\Delta \phi }}} \\ 
			{\left| {{e_2}} \right|} 
	\end{array}} \right]{e^{{i\phi _2}}},{\text{ }}\Delta \phi  = {\phi _1} - {\phi _2}.
\end{equation}
Typically, only the relative phase $\Delta \phi $ is relevant, and the global phase can be neglected. 

\indent When $\theta  = 0$, ${M_{{\text{MZI}}}}$ can be transformed into
\begin{equation}
{M_{{\rm{MZI}}}} = \left[ {\begin{array}{*{20}{c}}
		0&1\\
		1&0
\end{array}} \right] \cdot {e^{i{\pi  \mathord{\left/
				{\vphantom {\pi  2}} \right.
				\kern-\nulldelimiterspace} 2}}}.
\end{equation}
In this case, the 1-CPS is equivalent to the cross-connection structure (Fig. \ref{fig1} (b)).

\indent When $\theta  = \pi$, ${M_{{\text{MZI}}}}$ can be transformed into
\begin{equation}
{M_{{\rm{MZI}}}} = \left[ {\begin{array}{*{20}{c}}
		1&0\\
		0&1
\end{array}} \right] \cdot {M_{{\rm{PS}}}}\left( \pi  \right).
\end{equation}
In this case, the 1-CPS is equivalent to the direct-connection structure (Fig. \ref{fig1} (c)). 

\indent When $\theta  = {\pi  \mathord{\left/{\vphantom {\pi  2}}\right.\kern-\nulldelimiterspace} 2}$, ${M_{{\text{MZI}}}}$ can be transformed into
\begin{equation}
	\begin{aligned}
		M_{\mathrm{MZI}} =M_{\mathrm{ps}}(\pi / 2) \cdot M_{\mathrm{MMI}} \cdot M_{\mathrm{ps}}(\pi / 2) \cdot e^{i 7 \pi / 4} .
	\end{aligned}
\end{equation}
In this case, the 1-CPS is equivalent to a $2 \times 2$ MMI with two phase delays, ${\pi  \mathord{\left/{\vphantom {\pi  2}} \right.\kern-\nulldelimiterspace} 2}$, placed before and after it, respectively (Fig. \ref{fig1} (d)). 

\indent When $\theta  = {{3\pi } \mathord{\left/{\vphantom {{3\pi } 2}} \right.\kern-\nulldelimiterspace} 2}$, ${M_{{\text{MZI}}}}$ can be transformed into
\begin{equation}
	\begin{aligned}
		M_{\mathrm{MZI}} =M_{\mathrm{ps}}(3 \pi / 2) \cdot M_{\mathrm{MMI}} \cdot M_{\mathrm{ps}}(3 \pi / 2) \cdot e^{i \pi / 4} .
	\end{aligned}
\end{equation}
In this case, the 1-CPS is equivalent to a $2 \times 2$ MMI with two phase delays, ${{3\pi } \mathord{\left/{\vphantom {{3\pi } 2}} \right.\kern-\nulldelimiterspace} 2}$, placed before and after it (Fig. \ref{fig1} (e)).

\indent For any relative phase, $\theta $, when the input beam is ${e_{in}} = {\left[ {0,1} \right]^{\mathrm{T}}}$, the output beams are
\begin{equation}
	{e_{out}} = \left[ {\begin{array}{*{20}{c}}
			{{e_3}} \\ 
			{{e_4}} 
	\end{array}} \right] = {M_{{\text{MZI}}}} \cdot {e_{in}} = \frac{1}{2}\left[ {\begin{array}{*{20}{c}}
			{i\left( {1 + {e^{i\theta }}} \right)} \\ 
			{1 - {e^{i\theta }}} 
	\end{array}} \right].
\end{equation}
\indent Their corresponding intensities are 
\begin{equation}
	\begin{gathered}
		{I_3} = {\left| {{e_3}} \right|^2} = \frac{1}{2}\left[ {1 + \cos \left( \theta  \right)} \right], \hfill \\
		{I_4} = {\left| {{e_4}} \right|^2} = \frac{1}{2}\left[ {1 - \cos \left( \theta  \right)} \right]. \hfill \\ 
	\end{gathered} \label{Eq13}
\end{equation}

\begin{figure}[ph]
	\vspace{-0.7cm}
	\centering
	\includegraphics[scale=0.5]{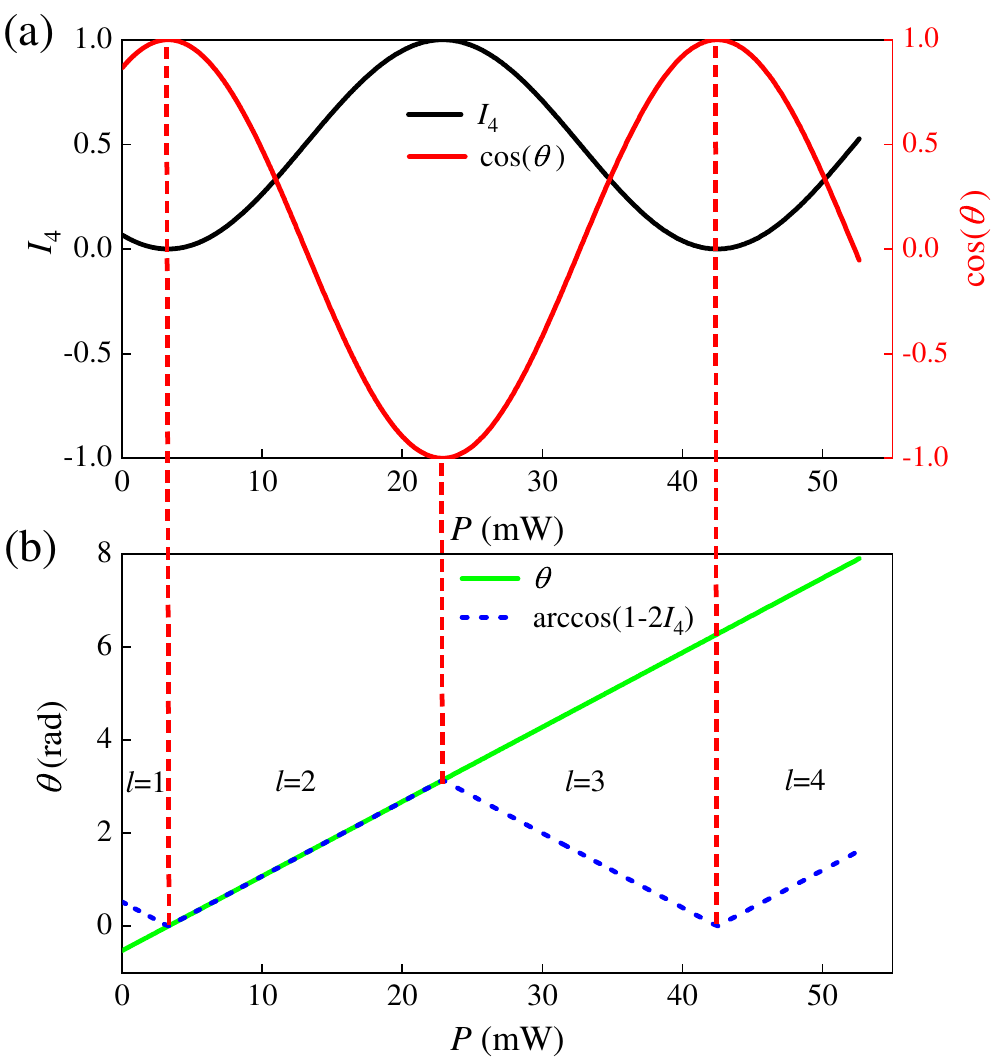}
	\caption{Calibration of the 1-CPS. (a) Intensity ${I_4}$ and $\cos \left( \theta  \right)$ versus the heating power $P$. (b) Relationship between the relative phase $\theta $ and heating power $P$.}
	\label{fig2}
\end{figure}

\indent During the calibration of 1-CPS, we only use ports 2 and 4. The laser beam injects into port 2, and outputs from port 4. The relationship between the intensity, ${I_4}$, and the applied heating power, $P$, on the TOPS is represented by the black curve shown in Fig. \ref{fig2} (a). Thus, the relationship between $P$ and $\cos \left( \theta  \right)$ can be derived using Eq. \ref{Eq13} (red curve, Fig. \ref{fig2} (a)). 

\indent By calculating the inverse function of $\cos \left( \theta  \right)$, blue dashed line were obtained (Fig. \ref{fig2} (b)). The linear relationship (green line, Fig. \ref{fig2} (b)) of relative phase $\theta $ and heating power $P$ is determined by correcting the blue dashed line using the following equation:
\begin{equation}
	\theta  = {\left( { - 1} \right)^l}\arccos \left( {1 - 2{I_4}} \right) + \left[ {l - \left( {1 + \frac{{1 + {{\left( { - 1} \right)}^l}}}{2}} \right)} \right]\pi ,{\text{ }}l = 1,2, ... \label{Eq14}
\end{equation}
where the positive integer, $l$, represents the $l{\text{th}}$ blue dashed line segments. By fitting the corrected line, we obtain the following linear equation:
\begin{equation}
	\theta  = kP + \Delta \theta, \label{Eq15}
\end{equation} 
where $k$ is the slope, and $\Delta \theta $ is the intercept. The constraint, $ |\Delta \theta|<\pi / 2$, is used to ensure the value of $l$.

\indent In the experiment, to ensure that the initial relative phase, $\left| {\Delta \theta } \right| < {\pi  \mathord{\left/{\vphantom {\pi  2}} \right.\kern-\nulldelimiterspace} 2}$, the lengths of both parallel waveguides should be short and equal during the design process. When the optical path of the upper waveguide is shorter, the initial relative phase $\Delta \theta  < 0$, and the initial value of $l$ is 1. Thus, it is necessary to compensate for a phase shift, ${\theta _{th}} < {\pi  \mathord{\left/{\vphantom {\pi  2}} \right.\kern-\nulldelimiterspace} 2}$, to ensure that the optical path is equal or that the relative phase is $\theta  = 0$. When the optical path of the upper waveguide is longer, the initial relative phase $\Delta \theta  > 0$, and the initial value of $l$ is 2. It is necessary to compensate a phase shift ${\theta _{th}} > {\pi  \mathord{\left/{\vphantom {\pi  2}} \right.\kern-\nulldelimiterspace} 2}$ to ensure that the optical path difference is half a wavelength or $\theta  = \pi $.      

\subsection{Calibration of 2-CPSs}
\begin{figure}[h]%
	\centering
	\includegraphics[scale=0.72]{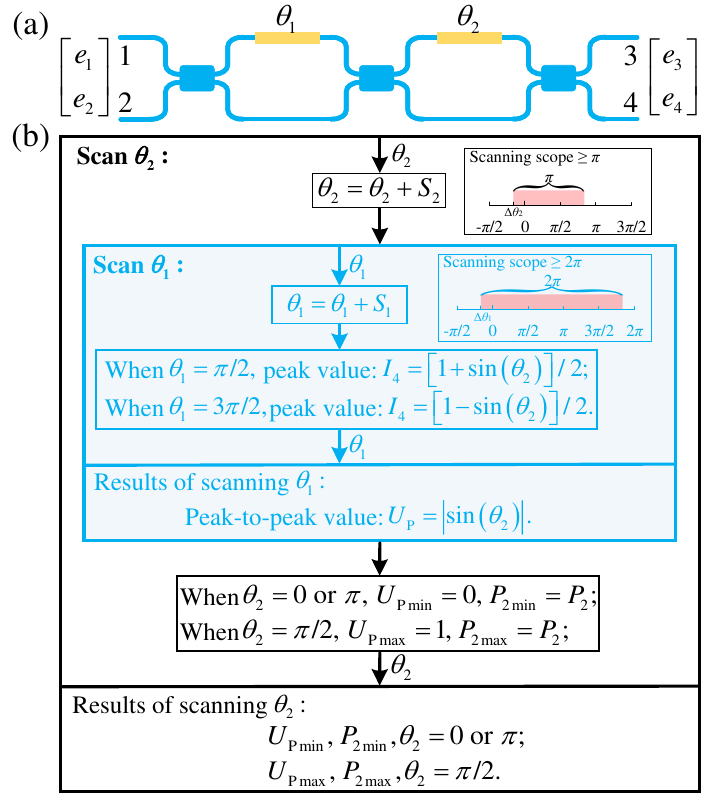} 
	\caption{Structure of 2-CPSs and the flowchart of pairwise scan method. (a) Structure of 2-CPSs. (b) The flowchart of pairwise scan method in calibrating 2-CPSs.} 
	\label{fig3}
\end{figure}

Fig. \ref{fig3} (a) shows the structure of 2-CPSs. The relative phases of the two pairs of waveguides are ${\theta _1}$ and ${\theta _2}$, respectively. Each relative phase has its initial relative and TOPS phases. They have a relationship similar to that in Eq. \ref{Eq3} as follows:
\begin{equation}
	{\theta _i} = {\theta _{thi}} + \Delta {\theta _i} = {k_i}{P_i} + \Delta {\theta _i},{\text{ }}i = 1,2. 
\end{equation} 

When the input beams are ${e_{in}} = {\left[ {0,1} \right]^{\mathrm{T}}}$, the output beams are
\begin{equation}
	\begin{aligned}
		e_{\text {out }} & =\left[\begin{array}{l}
			e_{3} \\
			e_{4}
		\end{array}\right]=M_{\mathrm{MMI}} \cdot M_{\theta_{2}} \cdot M_{\mathrm{MMI}} \cdot M_{\theta_{1}} \cdot M_{\mathrm{MMI}} \cdot e_{\text {in }} \\
		& =\frac{1}{2 \sqrt{2}} \cdot\left[\begin{array}{c}
			i\left(1-e^{i \theta_{1}}+e^{i \theta_{2}} \cdot\left(1+e^{i \theta_{1}}\right)\right) \\
			-1+e^{i \theta_{1}}+e^{i \theta_{2}} \cdot\left(1+e^{i \theta_{1}}\right)
		\end{array}\right].
	\end{aligned}
\end{equation}

Their corresponding intensities are
\begin{equation}
	\begin{gathered}
		{I_3} = \frac{1}{2}\left[ {1 - \sin \left( {{\theta _1}} \right)\sin \left( {{\theta _2}} \right)} \right], \hfill \\
		{I_4} = \frac{1}{2}\left[ {1 + \sin \left( {{\theta _1}} \right)\sin \left( {{\theta _2}} \right)} \right]. \hfill \\ 
	\end{gathered} \label{Eq17}
\end{equation}

\indent The pairwise scan method is employed for the calibration of 2-CPSs. The flowchart is shown in Fig. \ref{fig3} (b). When the laser beam transmits from ports 2 to 4, we scan the 2nd TOPS with step ${S_2}$ and a scanning scope $\ge \pi $. For each scanning step of the 2nd TOPS, we scan the 1st TOPS with step ${S_1}$ and a scanning scope $\ge 2\pi $. The peak-to-peak value output from the port 4 is denoted as ${U_{\rm{P}}} = \left| {\sin ({\theta _2})} \right|$.

\indent Based on Eq. \ref{Eq17} and the constraint $|\Delta \theta|<\pi / 2$, we can infer that the first minimum peak-to-peak value ${U_{{\rm{P}}\min}}$ corresponds to the relative phase ${\theta _2} = 0$ or ${\theta _2} = \pi $. The heating power applied to 2nd is denoted as ${P_{2\min}}$. The first maximun peak-to-peak value ${U_{{\rm{P}}\max}}$ corresponds to the relative phase $\theta _{2} = \pi / 2$. The heating power applied to 2nd is denoted as ${P_{2\max}}$. 

\indent When the heating power applied to the 2nd TOPS is ${P_{2\max }}$, we have $\sin \left( {{\theta _2}} \right) = \sin \left( {{\pi  \mathord{\left/{\vphantom {\pi  2}} \right.\kern-\nulldelimiterspace} 2}} \right) = 1$. Thus, Eq. \ref{Eq18} can be rewritten as 
\begin{equation}
	{I_4} = \frac{1}{2}\left[ {1 + \sin \left( {{\theta _1}} \right)} \right]. \label{Eq18}
\end{equation}

\begin{figure}[h]%
	\vspace{-0.3cm}
	\centering
	\includegraphics[scale=0.5]{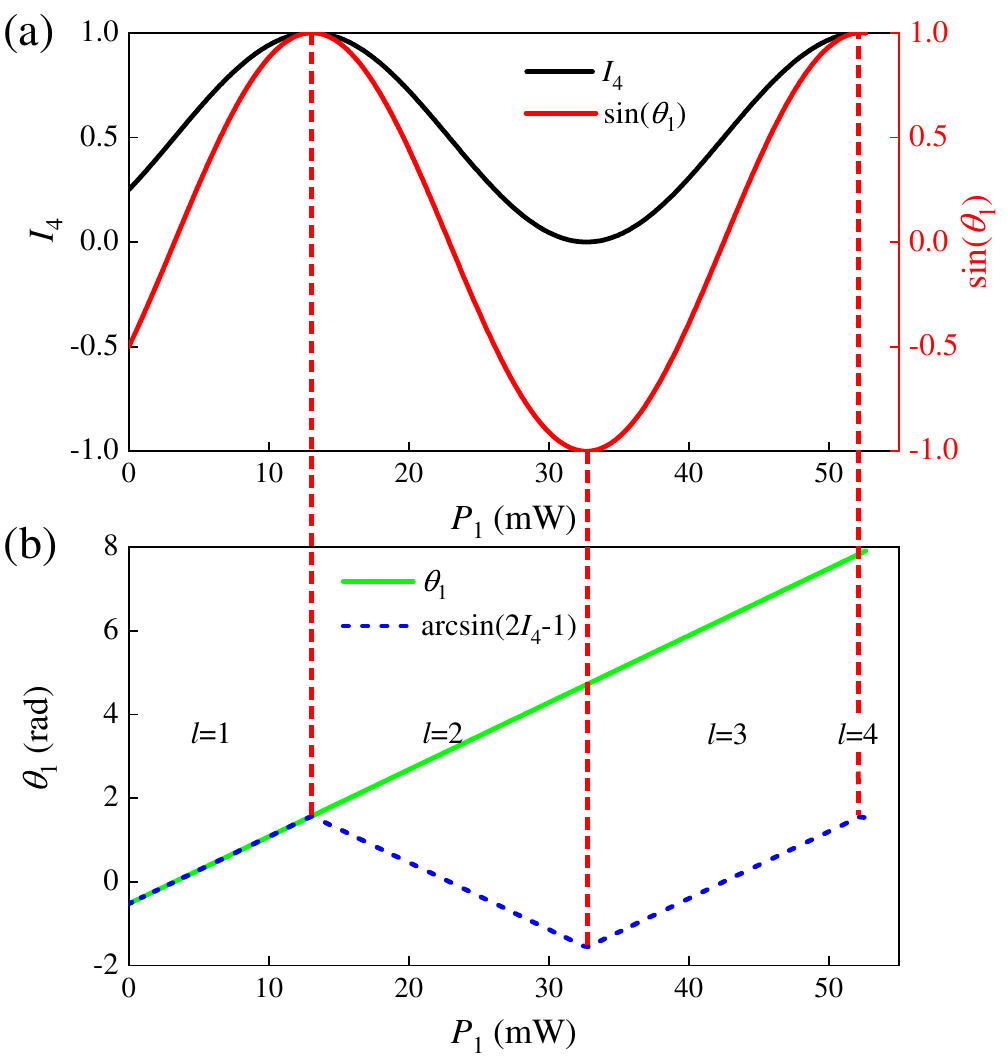}
	\caption{Calibration of the 2-CPSs. (a) Intensity ${I_4}$ and $\sin \left( {{\theta _1}} \right)$ versus heating power ${P_1}$. (b) Relationship between the relative phase ${\theta _1}$ and heating power ${P_1}$.} 
	\label{supfig4}
	\vspace{-0.6cm}
\end{figure}

\indent During the experiment, we determine the relationship between the intensity, ${I_4}$, and applied heating power, ${P_1}$ (black curve, Fig. \ref{supfig4} (a)). Using Eq. \ref{Eq18}, the relationship between $\sin \left( {{\theta _1}} \right)$ and ${P_1}$ can be determined (red curve, Fig. \ref{supfig4} (a)). By calculating the inverse function of $\sin \left( {{\theta}} \right)$, we obtain the blue dashed line (Fig. \ref{supfig4} (b)). The green linear relationship between ${\theta _1}$ and ${P_1}$ was determined by correcting the blue dashed line using the following equation:

\begin{equation}
	{\theta _1} = {\left( { - 1} \right)^{l - 1}}\arcsin \left( {2{I_4} - 1} \right) + \left( {l - 1} \right)\pi,
\end{equation}
where the positive integer, $l$, represents the $l{\text{th}}$ blue dashed line segments. By fitting the green line, we obtain the following linear equation: 
\begin{equation}
	{\theta _1} = {k_1}{P_1} + \Delta {\theta _1}.
\end{equation}

\indent The ${k_1}$ is the slope, and the $\Delta {\theta _1}$ is the intercept. Note that the constraint $\left| {\Delta \theta } \right| < {\pi  \mathord{\left/{\vphantom {\pi  2}} \right.\kern-\nulldelimiterspace} 2}$ should be used to ensure the value of $l$.

\begin{figure*}[bp]
	\vspace{-0.3cm}
	\centering
	\includegraphics[scale=0.7]{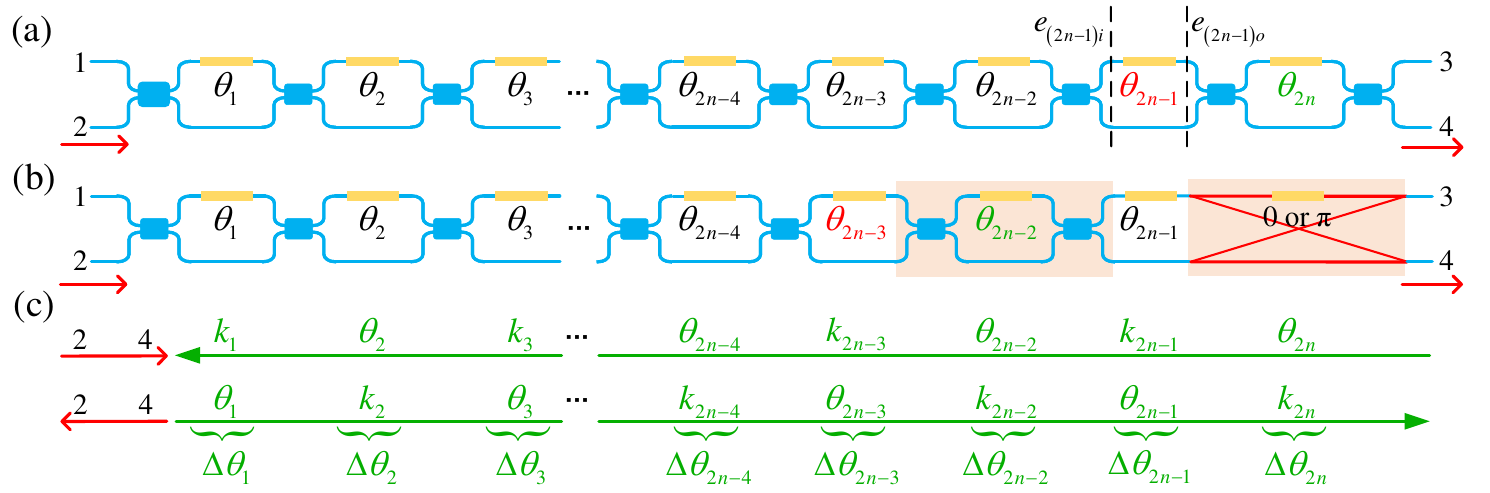}
	\caption{Structure of even CPSs and schematics of the calibration process. (a) Structure of even CPSs. (b) Equivalent structure of the $2n{\text{th}}$ TOPS in calibrating ${\theta _{2n - 2}}$, ${k_{2n - 3}}$. (c) Process of calibrating $k$, $\theta $, and $\Delta \theta $.} 
	\label{Figure5}
	\vspace{-0.3cm}
\end{figure*}

\indent By interchanging the scanning of 1st and 2nd TOPSs, a linear equation of the 2nd TOPS can be determined by the same way.

\subsection{Calibration method for even CPSs}
Fig. \ref{Figure5} (a) shows the structure of even CPSs. The total number of TOPSs is $N = 2n$, and the relative phase of the $j{\text{th}}$ phase shifter is   
\begin{equation}
	{\theta _j} = {\theta _{thj}} + \Delta {\theta _j} = {k_j}{P_j} + \Delta {\theta _j}, \label{Eq21}
\end{equation}
where $\Delta {\theta _j}$ is the initial relative phase of the $j{\text{th}}$ parallel waveguide, ${\theta _{thj}}$ is the phase generated by $j{\text{th}}$ TOPS, and ${P_j}$ is the heating power applied to the $j{\text{th}}$ TOPS.

\indent For the even CPSs, all the values of $k$ and $\Delta \theta $ should be calibrated. Using the pairwise scan method, we firstly calibrate the ${\theta _{2n}}$ and ${k_{2n - 1}}$ from the right side of the structure by transmitting the laser beam from ports 2 to 4. For clarity, the Jones vectors of laser beams prior to and after the $j{\text{th}}$ TOPS are denoted as ${e_{ji}} = {\left[ {{e_{j1}},{e_{j2}}} \right]^{\mathrm{T}}}$ and ${e_{jo}} = {\left[ {{e_{j3}},{e_{j4}}} \right]^{\mathrm{T}}}$, respectively. Then, the Jones vector before the $\left( {2n - 1} \right){\text{th}}$ TOPS is denoted as ${e_{\left( {2n - 1} \right)i}} = {\left[ {\left| {{e_{\left( {2n - 1} \right)1}}} \right|{e^{i\alpha }},\left| {{e_{\left( {2n - 1} \right)2}}} \right|} \right]^{\mathrm{T}}}$, $\alpha  \in \left[ {0,{\text{ 2}}\pi } \right)$, here $\alpha $ is the relative phase of the laser beams. Thus, the Jones vector of the laser beams output from ports 3 and 4 is
\begin{equation}
\begin{gathered}
	{e_{out}} = \left[ {\begin{array}{*{20}{c}}
			{{e_3}} \\ 
			{{e_4}} 
	\end{array}} \right] = {M_{{\text{MMI}}}} \cdot {M_{{\theta _{2n}}}} \cdot {M_{{\text{MMI}}}} \cdot {M_{{\theta _{2n - 1}}}} \cdot {e_{\left( {2n - 1} \right)i}} \hfill \\
	= \frac{1}{2}\left[ {\begin{array}{*{20}{c}}
			{\left| {{e_{\left( {2n - 1} \right)1}}} \right| \cdot \left( { - 1 + {e^{i{\theta _{2{n}}}}}} \right){e^{i\left( {\alpha  + {\theta _{2n - 1}}} \right)}} + i\left| {{e_{\left( {2n - 1} \right)2}}} \right| \cdot \left( {1 + {e^{i{\theta _{2n}}}}} \right)} \\ 
			{i\left| {{e_{\left( {2n - 1} \right)1}}} \right| \cdot \left( {1 + {e^{i{\theta _{2n}}}}} \right){e^{i\left( {\alpha  + {\theta _{2n - 1}}} \right)}} - \left| {{e_{\left( {2n - 1} \right)2}}} \right|\left( { - 1 + {e^{i{\theta _{2n}}}}} \right)} 
	\end{array}} \right]. \hfill \\ 
\end{gathered} 
\end{equation}

\indent The intensity of the laser beam output from port 4 is 
\begin{equation}
\begin{aligned}
	I_{4}= & \frac{1}{2}\left[1+\left(\left|e_{(2 n-1))_{1}}\right|^{2}-\left|e_{(2 n-1) 2}\right|^{2}\right) \cdot \cos \left(\theta_{2 n}\right)\right. \\
	& \left.-2\left|e_{(2 n-1) 1}\right| \cdot\left|e_{(2 n-1) 2}\right| \cdot \cos \left(\theta_{2 n-1}+\alpha\right) \cdot \sin \left(\theta_{2 n}\right)\right] \\
	= & \frac{1}{2}\left[1+c_{1} \cdot \cos \left(\theta_{2 n}\right)- c_{2} \cdot \cos \left(\theta_{2 n-1}+\alpha\right) \cdot \sin \left(\theta_{2 n}\right)\right] .
\end{aligned} \label{Eq24}
\end{equation}

\begin{figure}[h]%
	\centering
	\includegraphics[scale=0.72]{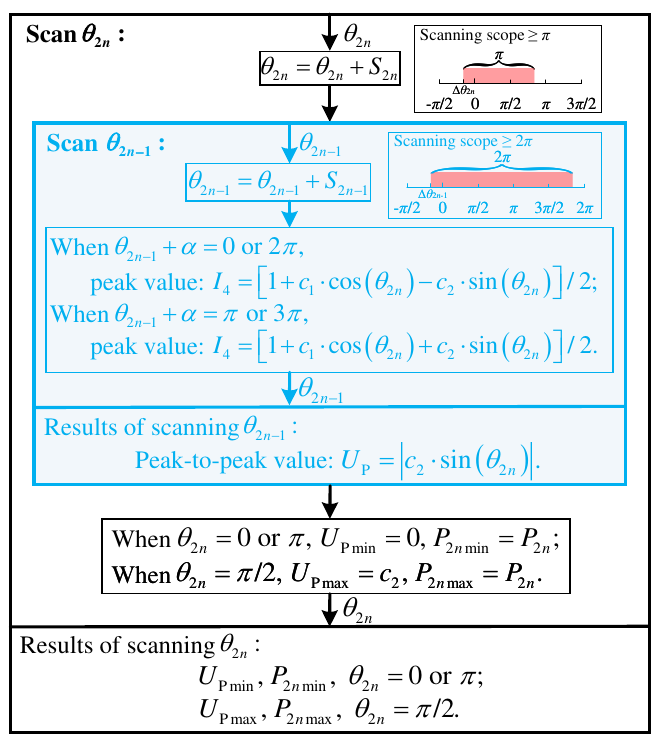}
	\caption{The flowchart of pairwise scan method in calibrating even CPSs.} 
	\label{supfig6}
\end{figure}

\indent When ${\theta _{2n}} = {\text{0}}$ or ${\theta _{2n}} = \pi $, ${I_4}$ remains constant, independent of the values of $\alpha $ and ${\theta _{2n - 1}}$. This conclusion can also be achieved from the equivalent structure as shown in Fig. \ref{Figure5} (b). When ${\theta _{2n}} = {\text{0}}$ or ${\theta _{2n}} = \pi $, the MZI structure with the $2n{\text{th}}$ TOPS is equivalent to the cross or direct-connection structure. In this case, the intensity, ${I_4}$, was independent of the value of $\alpha $ and ${\theta _{2n - 1}}$.

\indent We first scan the 2$n$th TOPS with step ${S_{2n}}$ and a scanning scope $\ge \pi$ as shown in Fig. \ref{supfig6}. For each scanning step of the 2$n$th TOPS, we performed a scan to the $\left( {2n - 1} \right){\text{th}}$ TOPS with step ${S_{2n - 1}}$ and a scanning scope $\ge 2\pi$. The peak-to-peak value of ${I_4}$ obtained by scanning the $\left( {2n - 1} \right){\text{th}}$ TOPS is denoted as ${U_{\rm{P}}} = \left| {{c_2} \cdot \sin \left( {{\theta _{2n}}} \right)} \right|$. The first minimum peak-to-peak value is denoted as ${U_{{\rm{P}}\min }} = 0$, and the heating power applied to the $2n{\text{th}}$ TOPS is denoted as ${P_{2n\min }}$. In this case, ${\theta _{2n}} = {\text{0}}$ or ${\theta _{2n}} = \pi $, and ${I_4} = c$. Combining the characters of Jones vectors and Eq. \ref{Eq24}, we can get
\begin{equation}
{c_2} = 2\sqrt {c\left( {1 - c} \right)} .
\end{equation}
 \begin{figure*}[bp]
	\vspace{-0.3cm}
	\centering
	\includegraphics[scale=0.7]{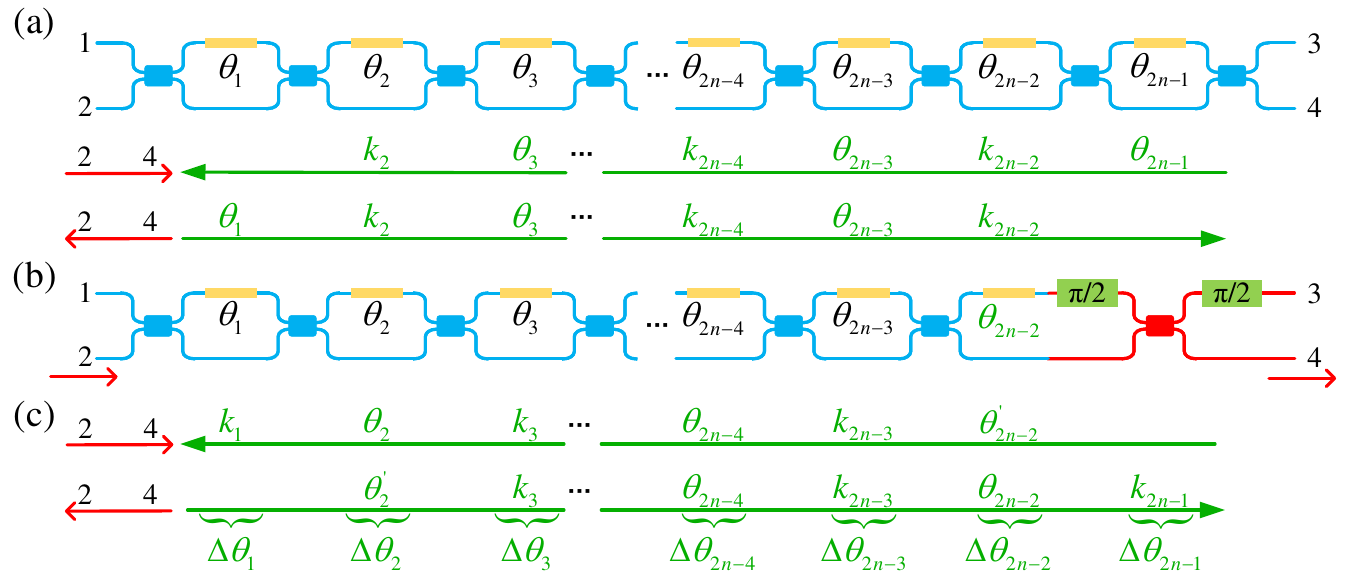}
	\caption{Structure of odd CPSs and schematics of the calibration process. (a) Structure of odd CPSs and the calibration scheme before using equivalent structure. (b) Equivalent structure of ($2n-1$)th TOPS. (c) Calibration schematics after using equivalent structure.}
	\label{Fig7}
	\vspace{-0.3cm}
\end{figure*}
\indent The first maximum peak-to-peak value is denoted as ${U_{{\rm{Pmax}}}} = {c_2}$, and the heating power applied to the $2n{\text{th}}$ TOPS is denoted as ${P_{2n\max }}$. In this case, ${\theta _{2n}} = \pi /2$. Eq. (\ref{Eq24}) is simplified to
\begin{equation}
	{I_4} = \frac{1}{2}\left[ {1 - {c_2} \cdot \cos \left( {\alpha  + {\theta _{2n - 1}}} \right)} \right].
\end{equation}
By calculating the inverse cosine function similar to Eq. \ref{Eq14}, we obtain the following linear equation:
\begin{equation}
	{\theta _{2n - 1}} = {k_{2n - 1}} \cdot {P_{2n - 1}} + \Delta {\theta _{2n - 1}} - \alpha. \label{Eq28}
\end{equation}

\indent Thus, the calibration of ${k_{2n - 1}}$ was completed. Owing to the relative phase of $\alpha $, $\Delta {\theta _{2n - 1}}$ could not be calibrated using Eq. \ref{Eq28}. However, we obtain ${P_{2n\min }}$ when ${\theta _{2n}} = {\text{0}}$ or ${\theta _{2n}} = \pi $, or we can say that we calibrate the relative phase, ${\theta _{2n}}$. Next, this result is subsequently used to calibrate $\Delta {\theta _{2n}}$.

\indent After the aforementioned steps, we apply ${P_{2n\min }}$ to the $2n{\text{th}}$ TOPS. In this case, the $2n{\text{th}}$ MZI structure is equivalent to the cross or direct-connection structure. Thereafter, we pairwise-scan and calibrate the $\left( {2n - 2} \right){\text{th}}$ and $\left( {2n - 3} \right){\text{th}}$ TOPSs as the $2n{\text{th}}$ and $\left( {2n - 1} \right){\text{th}}$ TOPSs.

\indent Using the pairwise scan methods above from right to left in a stepwise manner, we calibrate the relative phase, $\theta $, of all the even-number TOPSs and the slope, $k$, of all the odd-number TOPSs (Fig. \ref{Figure5} (c)).

\indent Thereafter, we change the transmission of laser beams from right to left. Using the pairwise scan methods above stepwise from left to right, we calibrate the $\theta $ of all the odd-number TOPSs and the $k$ of all the even-number TOPSs (Fig. \ref{Figure5} (c)).

\indent For any $j{\text{th}}$ TOPS, the applied heating power, ${P_{j\min }}$, corresponding to ${\theta _j} = 0$ or ${\theta _j} = \pi $ is calibrated, and the thermal relative phase ${\theta _{thj}} = {k_j}{P_j}$ could be calculated. Combining constraint $\left| {\Delta \theta } \right| < \pi /2$ and Eq. \ref{Eq21}, we calibrate the initial relative phase, $\Delta {\theta _j}$. Finally, we calibrate all the TOPSs in the even CPSs.

\indent In the even CPSs calibration, the constraint $\left| {\Delta \theta } \right| < \pi /2$ is used to simplify the calibration. Even if no constraint exists, the initial relative phase, $\Delta \theta $, is calibrated from 2nd to ($2n - 1$)th TOPSs. Details regarding the non-constraint even CPSs calibration method are provided in Appendix \ref{secA}.

\subsection{Calibration odd CPSs}
Fig. \ref{Fig7} shows the structure of odd CPSs, and the total number of TOPSs is $N = 2n - 1$. According to the calibration method of the even CPSs, we calibrate the relative phase, $\theta $, of the odd-number TOPSs ($2n - 1,2n - 3,...,5,3$) and the slope, $k$, of the even-number TOPSs ($2n - 2,2n - 4,...,4,2$). However, from left to right, the calibrated items are also the $ \theta $ ($1,3,...,2n - 3$) of the odd-number TOPSs and $k$ ($2,4, ...,2n - 4,2n-2$) of the even-number TOPSs (Fig. \ref{Fig7} (a)). Thus, to calibrate the $ \theta $ of the even-number TOPSs and the $k$ of the odd-number TOPSs, we introduce an equivalent structure of the $\left( {2n - 1} \right){\text{th}}$ TOPS (Fig. \ref{Fig7} (b)). In this case, the heating power, ${P_{\left( {2n - 1} \right)\max }}$, is applied to the $\left( {2n - 1} \right){\text{th}}$ TOPS to obtain ${\theta _{2n - 1}} = {\pi  \mathord{\left/{\vphantom {\pi  2}} \right.
\kern-\nulldelimiterspace} 2}$. Thereafter, we calibrate the $ \theta $ ($2n - 2,2n - 4,...,2$) of the even-number TOPSs and the $k$ ($2n - 3,...,3,1$) of the odd-number TOPSs from right to left when the laser beam transmits from ports 2 to 4 (Fig. \ref{Fig7} (c)). Using a similar method, when the laser beam transmits from ports 4 to 2, we can calibrate the $ \theta $ ($2,...,2n - 4,2n - 2$) of the even-number TOPSs and the $k$ ($3,...,2n-3,2n-1)$ of the odd-number TOPSs. Combining the calibration results shown in Fig. \ref{Fig7} (a) and (c), we can calibrate all the $k$ and $\Delta {\theta}$ of odd CPSs.

\indent The equivalent transform of the $\left( {2n - 1} \right){\text{th}}$ MZI structure introduces a ${\pi  \mathord{\left/{\vphantom {\pi  2}} \right.\kern-\nulldelimiterspace} 2}$ phase shift. When we calibrate the CPSs from right to left, the relative phase, $\theta _{2n - 2}^{'} = {\theta _{2n - 2}} + {\pi  \mathord{\left/{\vphantom {\pi  2}} \right.\kern-\nulldelimiterspace} 2}$, should be 0 or $\pi $. $\theta _2^{'}$ was in a similar situation.

\indent Using the odd CPSs calibration method, constraint $\left| {\Delta \theta } \right| < \pi /2$ is used to simplify the calibration. Even if no constraint existed, the initial relative phase, $\Delta \theta $, is calibrated from 2nd to ($2n - 2$)th TOPSs. Details of the non-constraint odd CPSs calibration method are available in Appendix \ref{secB}. 
\begin{figure*}[bp]
	\centering
	\includegraphics[scale=0.46]{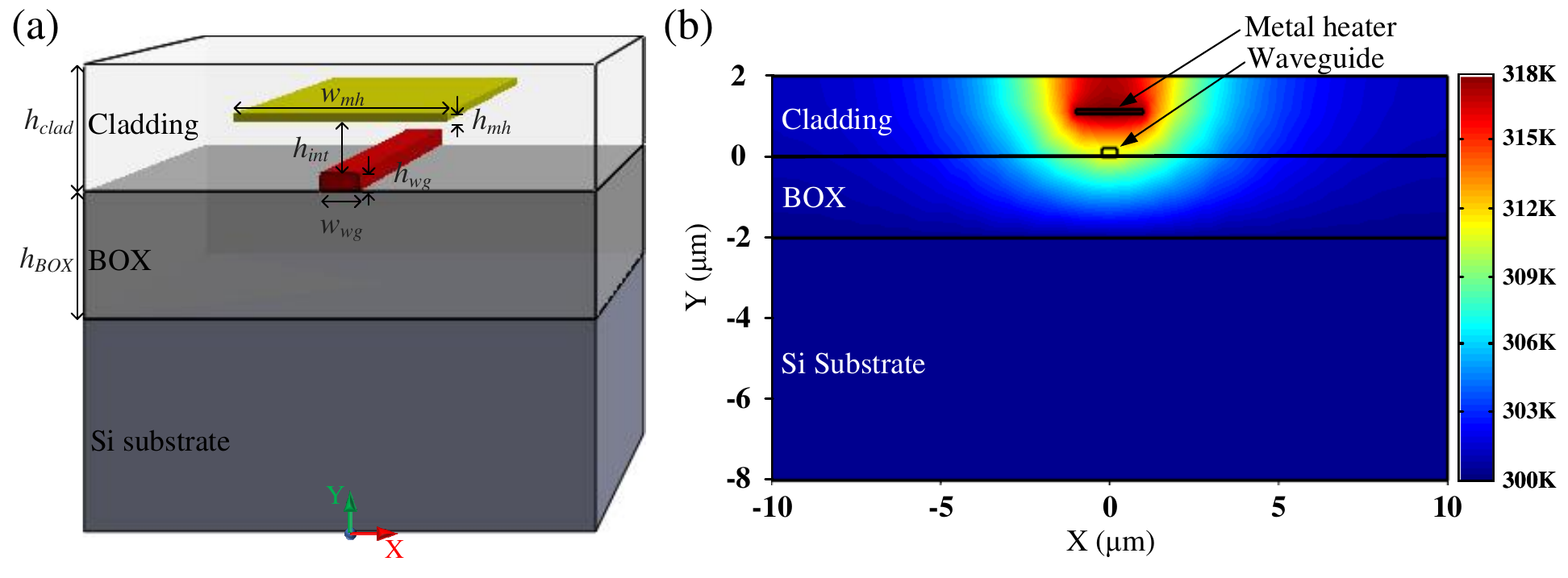}
	\caption{Structure of TOPS and temperature distribution in the silicon photonics chip. (a) Structure of TOPS on SOI platform. (b) Cross-section of temperature distribution in silicon photonics chip. } 
	\label{Fig8}
\end{figure*}
\section{Optimization of key CPS components} \label{sec:3}
The TOPS and $2 \times 2$ 50/50 MMI are key components of the CPS. The thermal crosstalk and phase-power linearity of the TOPS, and the balance of the $2 \times 2$ 50/50 MMI are closely related to the precision of calibration. To realize better calibration results, we simulate and optimize the structures of the key components.
\subsection{Simulation of TOPS}

The principle of TOPS is based on the thermal optical effect of silicon, which implies that effective refractive index, $n_{neff}$, of silicon varies with temperature $T$ as follows:
\begin{equation}
	\begin{gathered}
		{C_{TOPS}} = \frac{{d{n_{eff}}}}{{dT}} \hfill \\
		= 9.45 \times {10^{ - 5}} + 3.47 \times {10^{ - 7}} \times T - 1.49 \times {10^{ - 10}} \times {T^2}\left( {{{\text{K}}^{ - 1}}} \right){\text{,}} \hfill \\ 
	\end{gathered} \label{Eq29}
\end{equation}
where ${C_{TOPS}}$ is the thermal optical coefficient of silicon in the C band over the temperature range of 300--600 K \cite{ref26}. The linear term plays a dominant role within this range. Fig. \ref{Fig8} (a) shows the TOPS structure on the SOI platform. It comprises a silicon waveguide (red) and metal heater (yellow) above it. The metal heater transforms the electric energy into thermal energy, 
\begin{equation}
	P = {{{V^2}} \mathord{\left/
			{\vphantom {{{V^2}} R}} \right.
			\kern-\nulldelimiterspace} R},
\end{equation}
where $P$ is the heating power of TOPS, $V$ is the voltage applied to the metal heater, and $R$ is the resistance of the metal heater. The heating power changes the temperature of waveguide and the effective refractive index, ${n_{eff}}$. During the simulation, the heating power is set to set to 20 mW. The thicknesses of the buried oxide (BOX) layer, ${h_{BOX}}$, and cladding layer, ${h_{clad}}$, are both 2 $\upmu$m, and the thickness of the Si substrate is 750 $\upmu$m. The silicon waveguide is on the bottom of the cladding layer (width, ${w_{wg}} = 0.45$ $\upmu$m; height, ${h_{wg}} = 0.22$ $\upmu$m). The metal heater is approximately in the middle of the cladding layer (width, ${w_{mh}} = 2.5$ $\upmu$m; height, ${h_{mh}} = 0.1$ $\upmu$m). The interval between the metal heater and the waveguide is ${h_{int}} = 0.85$ $\upmu$m. The materials of BOX and cladding layer are  \ce{SiO2}, and that of the metal heater is Titanium nitride (TiN).

\indent The temperature distribution of the SOI chip is simulated using the Heat Transport (HEAT) solver of Ansys Lumerical by solving the following heat conduction equation:
\begin{equation}
	\rho {c_p}\frac{{\partial T}}{{\partial t}} - \nabla  \cdot \left( {{k_{hc}}\nabla T} \right) = Q,
\end{equation}
where $\rho $ is the mass density ($\mathrm{kg/m^{3} }$), ${c_p}$ is the specific heat ($\mathrm{J/(kg\cdot K)} $), ${k_{hc}}$ is the heat conductivity ($\mathrm{W/(m\cdot K)}$), and $Q$ is the heat energy transfer rate ($\mathrm{W/m^{3} }$) \cite{ref27}. $Q$ is calculated by
\begin{equation}
	Q = {P \mathord{\left/
			{\vphantom {P {\left( {{l_{mh}}{h_{mh}}{w_{mh}}} \right)}}} \right.
			\kern-\nulldelimiterspace} {\left( {{l_{mh}}{h_{mh}}{w_{mh}}} \right)}},
\end{equation}
where ${l_{mh}} = 390$ $\upmu$m is the length of the metal heater. The length satisfies ${l_{mh}} \gg {h_{mh}}$ and ${l_{mh}} \gg {w_{mh}}$; thus, the structure was assumed to exhibit translational symmetry, and Fig. \ref{Fig8} (b) shows the cross-section of the temperature distribution.
\begin{figure*}[tbph]%
	\centering
	\includegraphics[scale=0.75]{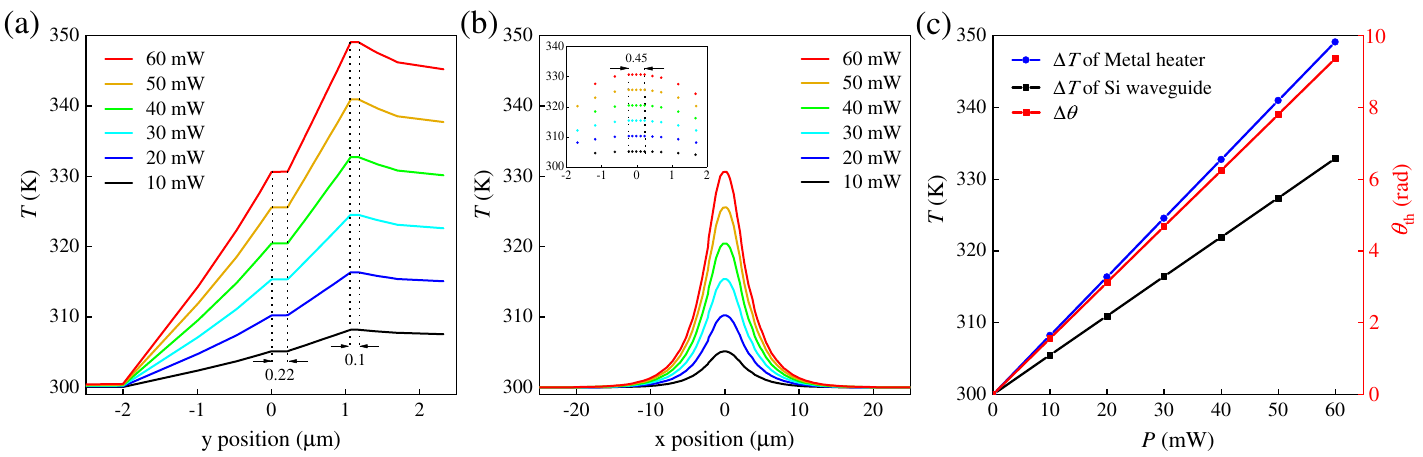}
	\caption{Temperature distribution of TOPS at different applied heating powers. (a) Temperature distribution in the $y$ direction at $x = 0$ $\upmu$m. (b) Temperature distribution in the $x$ direction at $y = 0.11$ $\upmu$m. (c) Temperature and relative phase ${\theta _{{\text{th}}}}$ versus different applied heating powers.} 
	\label{Fig9}
    \vspace{-0.3cm}
\end{figure*}

\indent The top of the device was bounded by air. During the simulation, the heat convection boundary condition, $- k{{\partial T} \mathord{\left/{\vphantom {{\partial T} {\partial n}}} \right.\kern-\nulldelimiterspace} {\partial n}} = {h_{air}}$, was applied to the top surface of the cladding, where $n$ was the direction normal to the surface and ${h_{air}} = 10$ $\mathrm{(W/(m^{2} \cdot K))} $ was the convection heat coefficient of the surrounding air. The Dirichlet boundary condition was applied to the bottom surface of the Si substrate. The temperatures of the top and bottom boundaries were set to ${T_{bc}} = 300$ K. The left and right boundaries of the simulation region were 25 $\upmu$m away from the TOPS structure such that there was practically no temperature variance at these boundaries. It is reasonable to set the boundary condition as thermal isolation. Table \ref{tab1} lists the parameters of various materials used in the simulation. 

\begin{table}[tbph]
	\vspace{-0.3cm}
	\begin{center}
		\caption{Parameters of various materials.} 	\label{tab1}
		\renewcommand{\arraystretch}{1.3}
		\tabcolsep 12pt 
		\begin{tabular}{| c | c | c | c |}
			\hline
			\multirow{2}{*}{Material} & $\rho $ & ${c_p}$ & ${k_{hc}}$ \\
               {} & ($\mathrm{kg/m^{3} }$) & ($\mathrm{J/(kg\cdot K)} $) & ($\mathrm{W/(m\cdot K)}$) \\\hline
               Si & 2330 & 711 & 148 \\
               \hline
        \ce{SiO2} & 2203 & 709 & 1.38 \\
               \hline
              TiN & 5430 & 604.45 & 67.7\\
               \hline
              air & 1.17 & 1006.43 & 0.026\\
			   \hline 
		\end{tabular}
	\end{center}
\vspace{-0.3cm}
\end{table}

\indent When we changed the heating power of the heater, the temperature distribution curves in $y$ direction at $x = 0$ $\upmu$m are shown in Fig. \ref{Fig9} (a). Considering that the heater conductivity, ${k_{hc}}$, of silicon and the metal heater considerably exceeded that of the surrounding \ce{SiO2}, the temperature in the waveguide and metal heater was uniform. Similarly, the temperature of the Si substrate was also uniform and close to the ${T_{bc}}$. The temperature distribution curves in the $x$ direction at $y = 0.11$ $\upmu$m with different heating power $P$ are shown in Fig. \ref{Fig9} (b). The peak value of each curve represents the temperature of the silicon waveguide. A good linear relationship exists between the temperature (black line, Fig. \ref{Fig9} (c)) and the heating power applied to the TOPS. Similarly, the linear blue line represents the temperature of the metal heater versus the heating power.

\indent The maximum applied voltage in the experiment was 10 V, and the resistance of the metal heater was approximately 1.75 k$\Omega $. Thus, the maximum heating power applied to the TOPS was 57 mW. The maximum heating power of 60 mW was adopted during the simulation. The simulation results showed that when the center distance of the TOPSs exceeded 40 $\upmu$m, nearly no thermal crosstalk was observed under the condition that the temperature was controlled. In addition, we have also performed a similar simulation along the direction of waveguide. Even if the heat conduction of Si is high, the simulation show that the temperature is close to the boundary temperature (300 K) when the length of waveguide not below the heater is above 20 $\upmu$m. Thus thermal crosstalk between CPSs is negligible considering the distance of 23 $\upmu$m between the heater and the MMI and waveguide length is 71 $\upmu$m.
 
\indent We used the finite element eigenmode solver to calculate the phase shift for different heating powers. The phase shift, ${\theta _{th}}$, of a TOPS is dependent on the change in the effective refractive index, ${n_{eff}}$, and length, ${l_{wg}} = {\text{390}}$ $\upmu$m, of the waveguide. It can be expressed as
\begin{equation}
	{\theta _{th}} = \frac{{2\pi }}{\lambda }\Delta {n_{eff}}{l_{wg}},
\end{equation}
where $\lambda  = {\text{1}}{\text{.55}}$ $\upmu$m represents the wavelength of the laser, and $\Delta {n_{eff}}$ can be calculated using Eq. \ref{Eq29} \cite{ref28}. Based on the simulation results (black line, Fig. \ref{Fig9} (c)), the relationship between the phase shift and heating power is shown as the red line with a slope of $ 0.1555$ rad/mW. It slightly exceeded the experimental result ($0.1477$ rad/mW). The main reason is that the metal (Al) used to connect the metal heater (TiN) absorbed part of heat quantity in a realistic situation.  

\indent The simulation revealed the temperature distribution of the TOPS on SOI platform. Importantly, a good linear relationship was observed between the phase shift ${\theta _{th}}$ of the TOPS and the applied power. The consumed heating power of the phase shift, $\pi $, was approximately 20.2 mW. Note that, we can infer the safe distance to prevent the thermal crosstalk of TOPSs.

\begin{figure*}[tbph]
	\centering
	\includegraphics[scale=0.6]{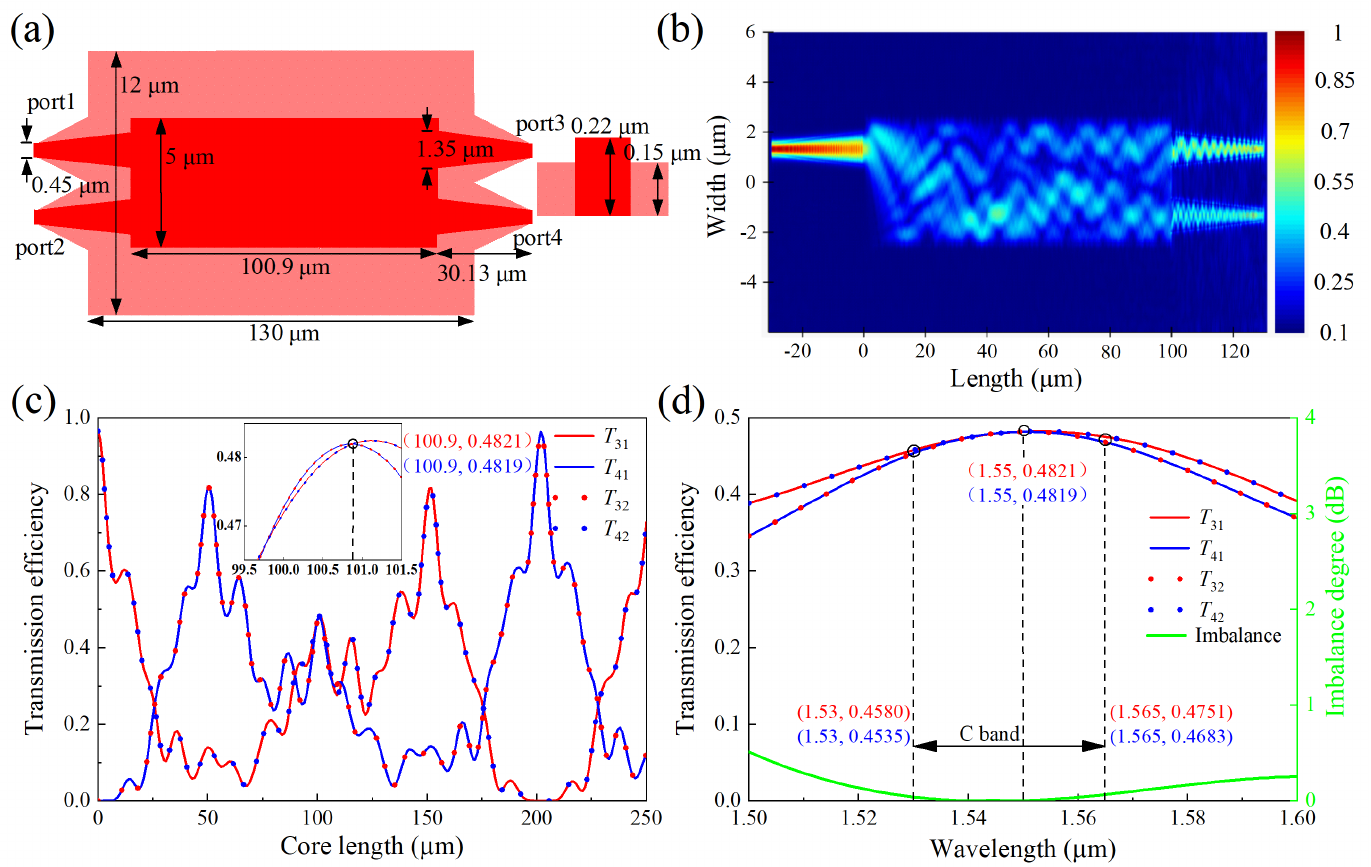}
	\caption{Optimized $2 \times 2$ 50/50 MMI structure. (a) Schematic of $2 \times 2$ 50/50 MMI structure. (b) Simulated optical power distribution at 1550 nm. (c) Transmission efficiency versus core length at 1550 nm. (d) Transmission efficiency and imbalance versus the wavelength.} 
	\label{Fig10}
    \vspace{-0.2cm}
\end{figure*}

\subsection{Simulation of $2 \times 2$ 50/50 MMI}
The balance degree of the $2 \times 2$ 50/50 MMI is a key parameter that directly affects the accuracy of the calibration results of CPS \cite{ref25}. Fig. \ref{Fig10} (a) shows an optimized structure of the $2 \times 2$ 50/50 MMI using the eigenmode expansion (EME) solver of Ansys Lumerical. The structure was fabricated based on a 70 nm-etched slab waveguide. Its core length and width were $L = 100.9$ $\upmu$m and $W = 5$ $\upmu$m, respectively. The two input ports were ports 1 and 2, and the two output ports were ports 3 and 4. The imbalance degree, $\Delta $ of two output intensities is usually defined as
 \begin{equation}
	\begin{gathered}
		{\Delta _1} = \left| {10 \cdot \log \left( {{{{T_{31}}} \mathord{\left/
						{\vphantom {{{T_{31}}} {{T_{41}}}}} \right.
						\kern-\nulldelimiterspace} {{T_{41}}}}} \right)} \right|\left( {\rm{dB}} \right), \hfill \\
		{\Delta _2} = \left| {10 \cdot \log \left( {{{{T_{32}}} \mathord{\left/
						{\vphantom {{{T_{32}}} {{T_{42}}}}} \right.
						\kern-\nulldelimiterspace} {{T_{42}}}}} \right)} \right|\left( {\rm{dB}} \right), \hfill \\ 
	\end{gathered} 
\end{equation}
where ${T_{31}}$ and ${T_{41}}$ are the transmission efficiencies from port 1 to ports 3 and 4, and ${T_{32}}$ and ${T_{42}}$ are the transmission efficiencies from port 2 to ports 3 and 4 \cite{ref29}. Fig. \ref{Fig10} (b) shows the top view of the simulated optical power distribution at 1550 nm. The source span was set to 2 $\upmu$m $\times$ 2 $\upmu$m during the simulation, and the laser beams were in the transverse electric (TE) mode in the structure.

\indent Fig. \ref{Fig10} (c) shows the transmission efficiency versus the core length when the wavelength of laser beam was 1550 nm. ${T_{31}}$ and ${T_{41}}$ were represented by the red and blue curves, respectively. ${T_{32}}$ and ${T_{42}}$ were represented by the red and blue dotted line, respectively. The sweep range of the core length, $L$, was 0--250 $\upmu$m with an interval of 1 $\upmu$m. The laser beams input from ports 1 and 2 exhibited good symmetrical characteristics. The splitting ratio varied with the $L$. We find that the 50/50 ratio corresponds to the core length of approximately 101 $\upmu$m. To simulate the balance degree with high precision, a small scan interval of 0.1 $\upmu$m was employed (inset in Fig. \ref{Fig10} (c)). The core length 100.9 $\upmu$m corresponds to a imbalance degree of approximately 0.0018 dB with ${T_{32}} = 0.4821$ and ${T_{42}} = 0.4819$. An extinction ratio of 73.5 dB can be achieved if the $2 \times 2$ MMI was used to construct the 1-CPS or MZI. Fig. \ref{Fig10} (d) shows the transmission efficiency and imbalance as a function of wavelength at the core length of 100.9 $\upmu$m. The red and blue curves represent ${T_{31}}$ and ${T_{41}}$, respectively. The red and blue dotted curves represent ${T_{32}}$ and ${T_{42}}$, respectively. The green curve shows that the imbalance degree varied with the wavelength. The simulated imbalance was lower than 0.063 dB in the range of C wave band (1530--1565 nm), and the extinction is higher than 42.8 dB. 

\indent In the experiment, the extinction ratio of several 1-CPSs were tested and were observed to exceed 50 dB, which was lower than the optimized value (73.5 dB) owing to the imperfections in fabrication. The 50 dB extinction ratio corresponds to an imbalance degree about 0.027 dB. With this imbalance degree, a calibration fidelity of 99.991$\%$ can be achieved. The reason can be seen in Appendix \ref{secC}. Thus, the realistic imbalance was sufficiently low to achieve a high fidelity. The method for calculating the fidelity is described in Section \ref{sec:4.3}. 

\section{Experimental calibration of 6-CPSs}
To verify the feasibility of the pairwise scan calibration method, a 6-CPSs was designed and fabricated using the industry-standard SOI technology of CUMEC \cite{Cumec}. The experimental setup and calibration results are as follows.

\subsection{Experimental setup}
\begin{figure*}[htbp]
	\centering
	\includegraphics[scale=0.6]{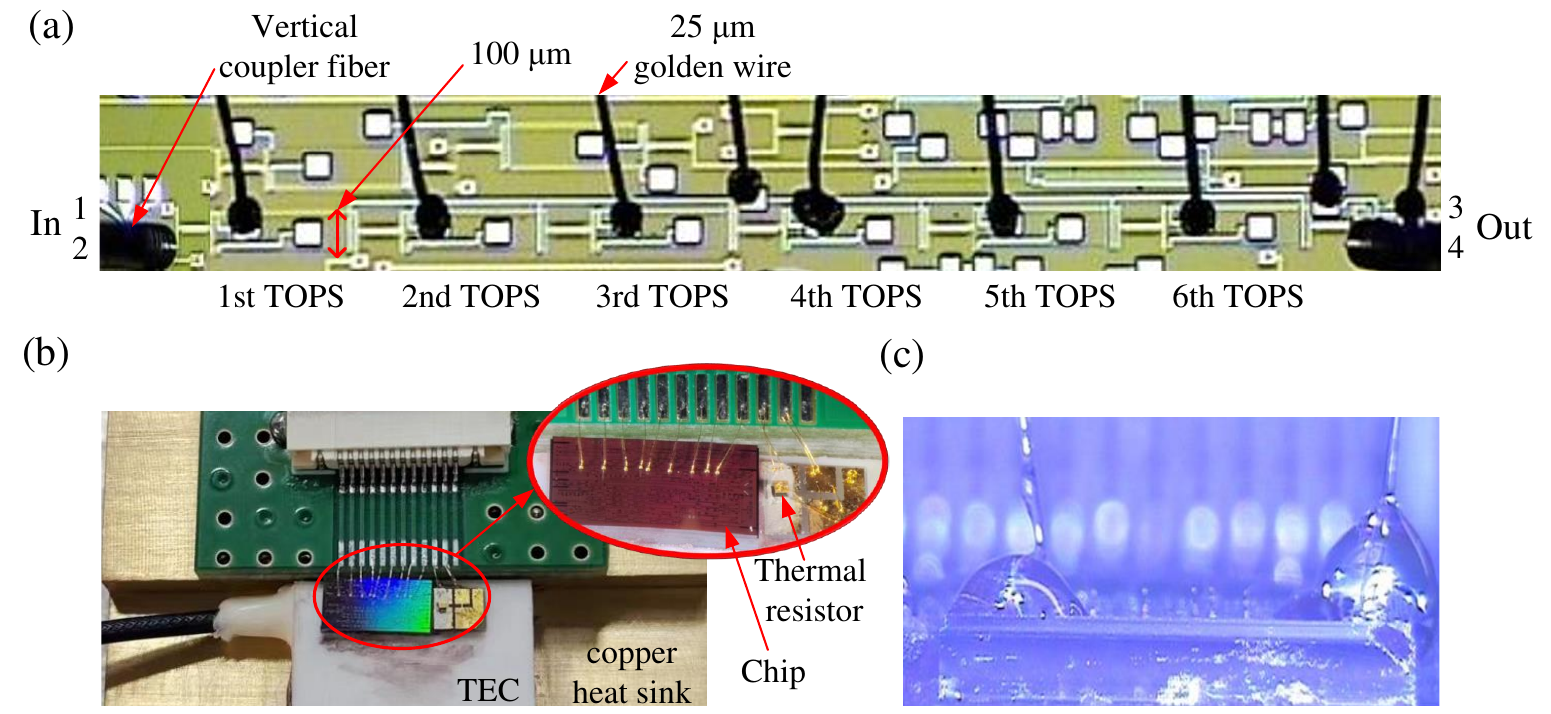}
	\caption{Microphotographs of the silicon photonics chip and experimental setup. (a) Microphotographs of 6-CPSs in the silicon photonics chip. (b) Microphotograph of the experimental setup. (c) Microphotograph of vertical coupling packaging.} 
	\label{Fig11}
	\vspace{-0.3cm}
\end{figure*}

Fig. \ref{Fig11} (a) shows the microphotograph of 6-CPSs in the silicon photonics chip. The distance of the upper and lower waveguides between the $2 \times 2$ MMIs is 100 $\upmu$m. The upper and lower waveguides both had TOPS for symmetry and backup. The upper TOPSs were used in the experiment. Ports 2 and 4 were one-dimensional (1D) grating couplers and were used to vertically couple the waveguides and single-mode fibers whose diameter was 125 $\upmu$m. Fig. \ref{Fig11} (b) shows the packaged chip. The heaters of the TOPSs were wire bonded to printed circuit board pads using 25 $\upmu$m golden wire. The entire silicon photonics chip adhered to the surface of the thermoelectric cooler (TEC) using thermal conductive adhesive. The thermal resistor was welded on a mount with good thermal conductivity. The mount was fixed on the TEC surface with thermal conductive adhesive and placed next to the chip to monitor its temperature. The TEC was mounted onto a copper heat sink. To enhance the calibration stability, ultraviolet (UV) glue was used in vertical coupling packaging (Fig. \ref{Fig11} (c)). The packaging increased the vertical coupling loss from 4.5 to 5.5 dB, and the total insertion loss of the chip was 12 dB.   

\subsection{Calibration results}
\begin{figure}[htbp] 
	\centering
	\includegraphics[scale=0.55]{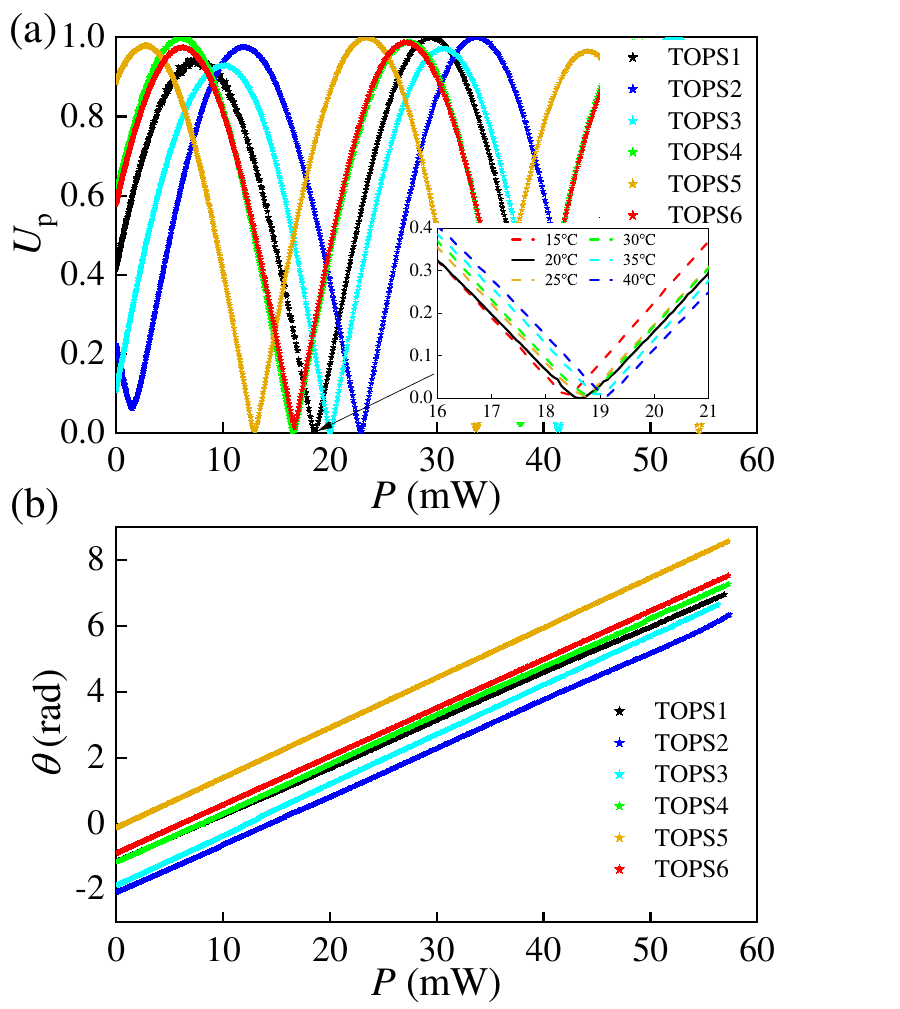}
	\caption{Experimental results for the 6-CPSs. (a) Curves of peak-to-peak value, ${U_p}$, versus the heating power $P$ (Inset figure: Curves about ${P_{1\min }}$ at different temperatures). (b) Linear relationship between the phase $\theta $ and the heating power $P$.} 
	\label{Fig12}
	\vspace{-0.5cm}
\end{figure}
During the experiment, two synchronized multifunction I/O cards (NI, USB6259) with six 16-bit digital-to-analog converters (DACs) were used to apply the voltage signals to the six TOPSs. A high resolution of 0.3 mV, which corresponds to a phase resolution of less than 0.028\degree or $4.9 \times {10^{ - 4}}$ radians, can be achieved when the output range is $ \pm $ 10 V \cite{ref30}. During the experimental scanning process,  the scanning range was 0--10 V, and the scanning step was 0.01 V corresponding to a phase resolution below 0.97\degree or $1.7 \times {10^{ - 2}}$ radians. The wavelength of the laser was 1550 nm, and the input optical power was 1 mW. The output laser was detected using a photodetector, and the output electrical signal was collected using an analog-to-digital converter (ADC; USB6259).

\indent During the CPSs calibration process, the temperature of the TEC was set to 20 $^{\circ}$C. When the laser beam transmitted from ports 2 to 4, we firstly calibrated the relative phase, ${\theta _6}$, of the 6th TOPS and the slope, ${k_5}$, of the 5th TOPS. For each scan step of the 6th TOPS, the peak-to-peak value obtained by scanning the 5th TOPS was recorded as ${U_{p6}}$. When the first pairwise scan was completed, the trace of the peak-to-peak value, ${U_{p6}}$, was plotted (red curve, Fig. \ref{Fig12} (a)). The first appearance of the maximum, ${U_{p6\max }}$, corresponded to ${\theta _6} = {\pi  \mathord{\left/{\vphantom {\pi  2}} \right.\kern-\nulldelimiterspace} 2}$, and its applied heating power was ${P_{6\max }}$. The first appearance of the minimum, ${U_{p6\min }}$, corresponded to ${\theta _6}$ with a value of 0 or $\pi$, and the applied heating power was ${P_{6\min }}$. In this case, the normalized intensity of the laser beam output from port 4 was a fixed value, ${c_6}$. When the heating power of the 6th TOPS was ${P_{6\max }}$, the orange line comprising asterisk points shown in Fig. \ref{Fig12} (b) was obtained by transforming the scan data of the 5th TOPS using the inverse cosine function. Through the linear fitting of the orange points, the slope, ${k_5}$, was obtained. To calibrate the 4th and 3rd TOPS, ${P_{6\min }}$ was applied to the 6th TOPS such that it was equivalent to a cross or direct-connection structure. 

\indent Fig. \ref{Fig12} (a) shows the trace of the peak-to-peak value, ${U_{p4}}$, during calibration as a green curve, and Fig. \ref{Fig12} (b) shows the cyan line of 3rd TOPS. Using a similar procedure, we obtained blue, black, cyan, and orange curves (Fig. \ref{Fig12} (a)) corresponding to the peak-to-peak values of ${U_{p2}}$, ${U_{p1}}$, ${U_{p3}}$, and ${U_{p5}}$, respectively. The black, blue, green, and red asterisks in Fig. \ref{Fig12} (b) represent the data of the 1st, 2nd, 4th and 6th TOPS, respectively. Afterward, the slopes (${k_1}$, ${k_2}$, ${k_4}$ and ${k_6}$) were obtained. Combining the $k$, ${P_{\min }}$, and constraint $\left| {\Delta \theta } \right| < {\pi  \mathord{\left/{\vphantom {\pi  2}} \right.\kern-\nulldelimiterspace} 2}$, the initial phase, $\Delta \theta $, is calculated by Eq. \ref{Eq15}. Table \ref{tab2} lists the calibration results for the 6-CPSs. The calibration results with constraint were consistent with the calibration results without constraint (Table \ref{tabA1}).  
\begin{table}[htbp]
	\vspace{-0.2cm}
	\begin{center}
		\caption{Calibration results for 6-CPSs.} 	\label{tab2}
		\renewcommand{\arraystretch}{1.3}
		\tabcolsep 3.5pt 
		\begin{tabular}{| c | c | c | c | c | c | c |}
			\hline
			\multirow{2}{*}{Result} & 1st & 2nd & 3rd & 4th & 5th & 6th\\
			{}  & TOPS & TOPS & TOPS & TOPS & TOPS & TOPS\\\hline
			${P_{\min }}$ (mW) & 18.6075 & 1.4921 & 20.0369 & 16.5703 & 12.9558 & 16.4435 \\
			\hline
			$k$ (rad/mW) & 0.1427 & 0.1459 & 0.1515 & 0.1478 & 0.1517 & 0.1470 \\
			\hline
			$\Delta \theta $ (rad) & 0.4863 & -0.2177 & 0.106 & 0.6925 & 1.1762 & 0.7244\\
			\hline
			$\Delta \theta $ (degree) & 27.86 & -12.47 & 6.07 & 39.68 & 67.39 & 41.5\\
			\hline 
		\end{tabular}
	\end{center}
    \vspace{-0.2cm}
\end{table} 

\begin{table*}[bp] 
	\vspace{-0.2cm}
	\footnotesize
	\caption{The calibration results of 1st TOPS at different temperatures} \label{tab3}
	\renewcommand{\arraystretch}{1.3}
	\tabcolsep 13pt 
	\begin{tabular*}{\textwidth}{cccccccccc}
		\hline	
		$T$ &Result &1th &2th &3th &4th &5th &6th &mean &deviation \\\hline
		{} &${k_1}$ &0.1428 & 0.1428 &0.1427 &0.1427 &0.1427 &0.1427 &0.1427 &$1.71 \times {10^{ - 5}}$ \\
		15 $^{\circ}$C &$\Delta {\theta _1}$ (rad) &0.5039 & 0.5130 &0.5043 &0.5041 &0.5047 &0.5138 &0.5073 &0.0043 \\
		{} &$\Delta {\theta _1}$ (degree) &28.87 & 29.39 &28.90 &28.88 &28.92 &29.44 &29.07 &0.2471 \\\hline
		
		{} &${k_1}$ &0.1427 & 0.1427 &0.1427 &0.1427 &0.1427 &0.1427 &0.1427 &$6.87 \times {10^{ - 6}}$ \\
		20 $^{\circ}$C &$\Delta {\theta _1}$ (rad) &0.4865 & 0.4954 &0.4770 &0.4863 &0.4861 &0.4770 &0.4847 &0.0063 \\
		{} &$\Delta {\theta _1}$ (degree) &27.86 & 28.38 &27.33 &27.86 &27.85 &27.33 &27.86 &0.3628\\\hline
		
		{} &${k_1}$ &0.1432 & 0.1430 &0.1428 &0.1427 &0.1431 &0.1428 &0.1429 &$1.81 \times {10^{ - 4}}$ \\
		25 $^{\circ}$C &$\Delta {\theta _1}$ (rad) &0.4583 & 0.4714 &0.4658 &0.4671 &0.4695 &0.4667 &0.4665 &0.0041 \\
		{} &$\Delta {\theta _1}$ (degree) &26.26 & 27.01 &26.69 &26.76 &26.90 &26.74 &26.73 &0.2354\\\hline
		
		{} &${k_1}$ &0.1428 & 0.1428 &0.1427 &0.1431 &0.1429 &0.1434 &0.1429 &$2.37 \times {10^{ - 4}}$ \\
		30 $^{\circ}$C &$\Delta {\theta _1}$ (rad) &0.4564 & 0.4574 &0.4578 &0.4414 &0.4452 &0.4452 &0.4506 &0.0068 \\
		{} &$\Delta {\theta _1}$ (degree) &26.15 & 26.21 &26.23 &25.29 &25.51 &25.51 &26.15 &0.3876\\\hline
		
		{} &${k_1}$ &0.1427 & 0.1429 &0.1432 &0.1428 &0.1429 &0.1428 &0.1429 &$1.57 \times {10^{ - 4}}$ \\
		35 $^{\circ}$C &$\Delta {\theta _1}$ (rad) &0.4490 & 0.4358 &0.4301 &0.4377 &0.4358 &0.4377 &0.4377 &0.0056 \\
		{} &$\Delta {\theta _1}$ (degree) &25.72 & 24.97 &24.64 &25.08 &24.97 &25.08 &25.08 &0.3237\\\hline
		
		{} &${k_1}$ &0.1435 & 0.1428 &0.1428 &0.1428 &0.1432 &0.1429 &0.1430 &$2.65 \times {10^{ - 4}}$ \\
		40 $^{\circ}$C &$\Delta {\theta _1}$ (rad) &0.4150 & 0.4189 &0.4189 &0.4283 &0.4113 &0.4075 &0.4166 &0.0066 \\
		{} &$\Delta {\theta _1}$ (degree) &23.78 & 24.00 &24.00 &24.54 &23.56 &23.35 &23.87 &0.3775\\\hline
	\end{tabular*}
\end{table*}

\indent Using the minimum slope ${k_1}$ and the maximun slop ${k_5}$, a phase shift difference of 0.36 rad will be generated by applying 40 mW heating power that can cause a phase shift about $2\pi $. It means that of fixed slope value is used for all TOPSs, an error with almost the same magnitude to the initial relative phase can be generated. Therefore the accurate calibration of each TOPS is necessary.

\indent Temperature is the major factor affecting the standrad deviation of the initial phase $\Delta {\theta}$. To observe the influence of temperature, we set the temperature of the chip to 15 $^{\circ}$C, 20 $^{\circ}$C, 25 $^{\circ}$C, 30 $^{\circ}$C, 35 $^{\circ}$C and 40 $^{\circ}$C using TEC and calibrate the 1st TOPS at corresponding temperature. At each temperature, the 1st TOPS was calibrated six times. Table \ref{tab3} presents the calibration results of the 1st TOPS.

\indent The calibration results at the same temperature exhibited good consistency. The slope, ${k_1}$, remained unchanged, and the maximum standard deviation was $2.73 \times {10^{ - 4}}$. The maximum standard deviation of the initial phase, $\Delta {\theta _1}$, was 0.39\degree. Comparing the results of different temperatures, we observed that $\Delta {\theta _1}$ between both arms gradually decreased with an increase in temperature. However, the slope remained unchanged. When the temperature was increased from 15 $^{\circ}$C to 40 $^{\circ}$C, the initial relative phase, $\Delta {\theta _1}$, decreased by 5.2\degree. The inset in Fig. \ref{Fig12} (a) shows ${P_{1\min }}$ during the calibration. In addition, the influence of temperature on the $\Delta {\theta _1}$ can also be inferred. ${P_{1\min }}$ gradually increased with an increase in temperature. 

\subsection{Fidelity testing} \label{sec:4.3} 
\indent The fidelity was used to verify the accuracy of our CPSs calibration method. The TEC temperature was set to 20 $^{\circ}$C, and a six-channel DAC was used to sequentially apply the scanning signals to the six TOPSs. The scanning range was 0--10 V with a scanning step of 0.01 V, which results in a 1000 scanning points for each TOPS. The expermental normalized intensity of the laser beam outputs from port 4 was defined as ${I_{4en}},\left( {n = 1,2, \cdots ,6000} \right)$. To calculate the fidelity, the expected output intensity ${I_{4tn}}$ of the laser from port 4 is derived based on the theoretical 6-CPSs model constructed using the calibration results. Thereafter, 6000 fidelity values were calculated using the following equation \cite{ref31}:
\begin{equation}
	{F_n} = \sqrt {{I_{4en}}{I_{4tn}}}  + \sqrt {\left( {1 - {I_{4en}}} \right)\left( {1 - {I_{4tn}}} \right)} .
\end{equation}

\begin{figure}[htbp]
    \vspace{-0.5cm}
	\centering
	\includegraphics[scale=0.35]{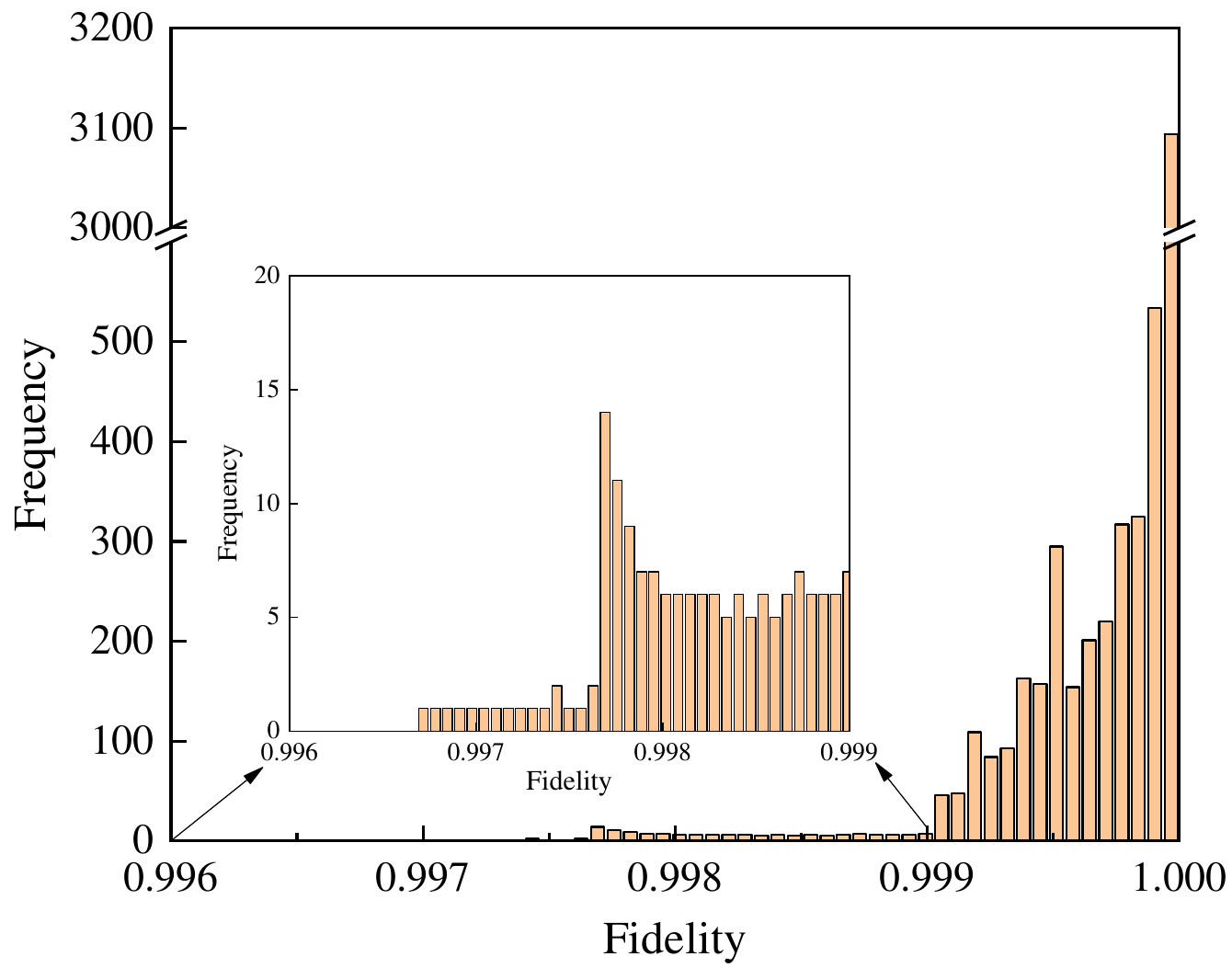}		
	\caption{The statistical results of the fidelity values.} 
	\label{fig13}
\end{figure}

\indent Fig. \ref{fig13} shows the statistical results of the fidelity. The average, maximum, and minimum values were 99.97$\%$, 100$\%$, and 99.68$\%$, respectively. In addition, 96.6$\%$ of the data had a fidelity above 99.9$\%$.   
\begin{figure*}[bp]
	\vspace{-0.2cm}
	\renewcommand{\thefigure}{A1}
	\centering
	\includegraphics[scale=0.65]{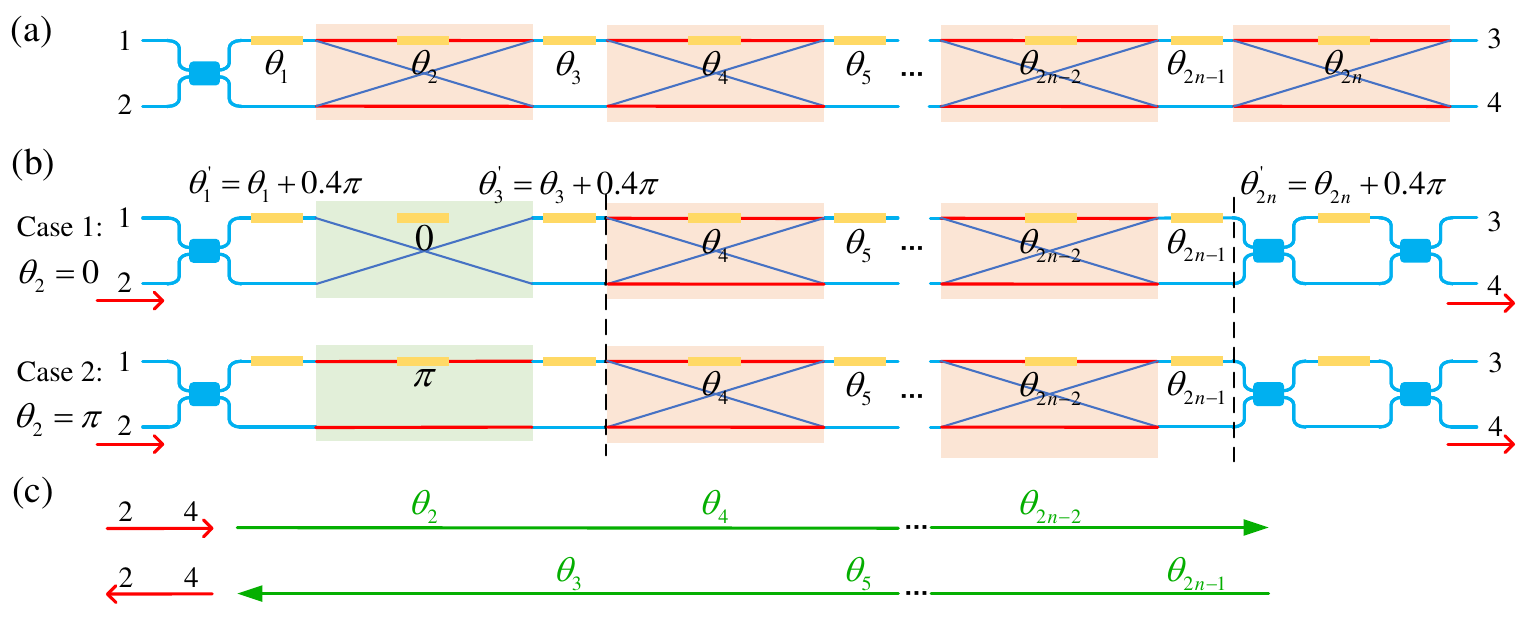}
	\caption{Schematics of calibrating $\theta $ in even CPSs without constraint. (a) Equivalent structure when the relative phases $\theta$ of even-number TOPSs were 0 or $\pi$. (b) Schematics of calibrating ${\theta _2}$ of 2nd TOPS. (c) Procedure of calibrating $\theta $ from 2nd to $\left( {2n - 1} \right)$th TOPSs.} 
	\label{A1}
	\vspace{-0.2cm}
\end{figure*}

\section{Conclusion}\label{sec:5}
This paper reported a pairwise scan method that can rapidly calibrate the CPSs structure in a silicon photonics integrated chip. Four kinds of equivalent structure of 1-CPS and reasonable constraint $\left| {\Delta \theta } \right| < {\pi  \mathord{\left/{\vphantom {\pi  2}} \right.\kern-\nulldelimiterspace} 2}$ were used to simplify the calibration. The calibration methods slightly differed according to the parity of the series number $N$ of CPSs. Using the pairwise scan method, only one input port and one output port are needed. This flexibility will make the calibration convenient when the CPSs are used to constitute more complex networks \cite{ref3}. In addition, calibration methods without constraint $\left| {\Delta \theta } \right| < {\pi  \mathord{\left/ {\vphantom {\pi  2}} \right.\kern-\nulldelimiterspace} 2}$ were introduced. After the scanning process, only a little amount of calculation is required to accomplish the calibration.     

\indent To reasonably design the CPSs structure, the key components, the TOPS and $2 \times 2$ 50/50 MMI, were simulated and optimized to prevent thermal crosstalk and obtain a better 50/50 splitting ratio. With the optimized $2 \times 2$ 50/50 MMI, the extinction ratio of 1-CPS can exceed 50 dB, which can ensure that the theoretical fidelity exceeds 99.991$\%$. A 6-CPSs structure in a packaged silicon photonics chip under different ambient temperatures was used to verify the rapid pairwise scan method. The calibration results with the constraint $\left| {\Delta \theta } \right| < \pi /2$ coincides with the case without the constraint, and a fidelity of 99.97$\%$ was achieved. Our results can find practical applications on large-scale quantum circuits in photonics integrated chip. In our future work, we will try to improve the pairwise scan method to calibrate the Resh and Clements structures efficiently \cite{ref44,ref45,ref46,ref47,ref48,ref49}.
\begin{figure*}[bp]
	\renewcommand{\thefigure}{B1}
	\centering
	\includegraphics[scale=0.65]{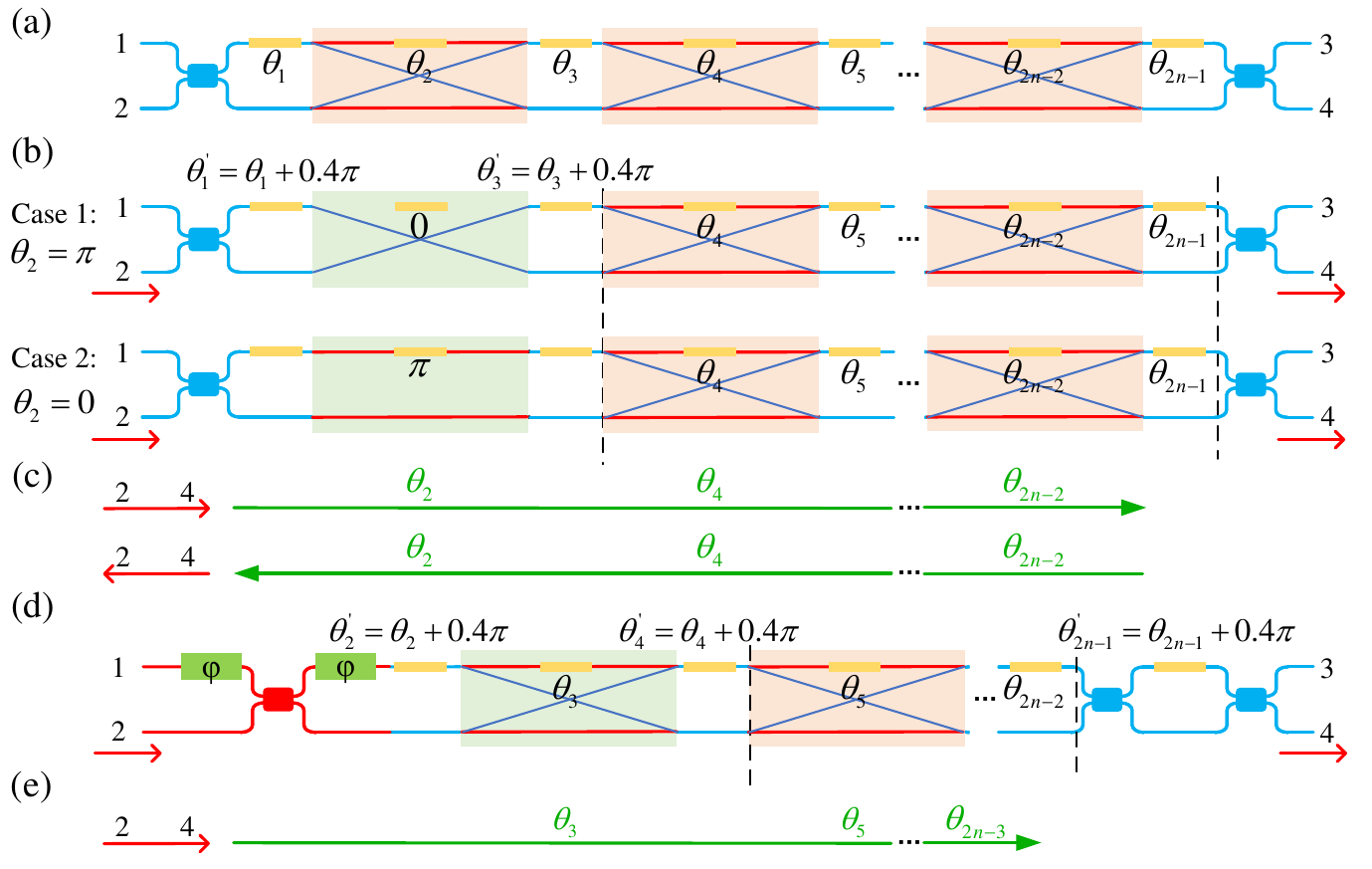}
	\caption{Schematics of calibrating $\theta $ in odd CPSs without constraint. (a) Equivalent structure when the relative phases $\theta$ of even-number TOPSs are 0 or $\pi$. (b) Schematics of calibrating ${\theta _2}$ of the 2nd TOPS. (c) Procedure of calibrating even-number TOPSs (2, 4, ..., $2n - 2$). (d) Schematics of calibrating ${\theta _3}$ of the 3rd TOPS. (e) Procedure of calibrating odd-number TOPSs (3, 5, ..., $2n - 3$).} 
	\label{B1}
\end{figure*}
\appendices
\section{Method of calibrating ${\theta _j}$ in even CPSs without constraint} \label{secA}
\setcounter{equation}{0}
\setcounter{figure}{0}
\setcounter{table}{0}
\renewcommand{\theequation}{A\arabic{equation}}
\renewcommand{\thefigure}{A\arabic{figure}}
\renewcommand{\thetable}{A\arabic{table}}

In the even CPSs, the calibration method introduced below did not require a constraint when calibrating $\Delta \theta $ from 2nd to $\left( {2n - 1} \right)$th TOPSs.

\indent The relative phase, $\theta $, of each TOPS can be calibrated to either 0 or $\pi$ using the following even CPSs calibration method without constraint $\left| {\Delta \theta } \right| < {\pi  \mathord{\left/{\vphantom {\pi  2}} \right.\kern-\nulldelimiterspace} 2}$. When all the relative phases were set to 0 or $\pi$, the MZI structures with the even-number TOPS are equivalent to the cross or direct-connection structures (Fig. \ref{A1} (a)). This was the initial state of the subsequent calibration, and the even CPSs were nearly equivalent to a $2 \times 2$ MMI. When the input beam was ${e_{in}} = {\left[ {0,1} \right]^{\mathrm{T}}}$, the output beam from the first $2 \times 2$ MMI becomes $e_{1 i}=\left[e^{i \pi / 2} / \sqrt{2}, 1 / \sqrt{2}\right]^{\mathrm{T}}$. Subsequently, when the beams pass through the cross or direct-connection structure, the intensity remained at 1/2 even if the relative phases of the TOPSs were unknown.

\indent To determine whether the relative phase ${\theta _2}$ was 0 or $\pi$, 0.4$\pi$ phase shifts were applied to the 1st and 3rd TOPS, which changed the relative phase to $\theta _1^{'} = {\theta _1} + 0.4{\pi}$ and $\theta _3^{'} = {\theta _3} + 0.4{\pi}$, respectively. When ${\theta _2} = 0$, the MZI structure with the 2nd TOPS was equivalent to the cross structure (Fig. \ref {A1} (b)). The two 0.4$\pi$ phase shifts applied to the 1st and 3rd TOPS act on different paths, which led to no relative phase change. When ${\theta _2} = {\pi}$, the MZI structure with 2nd TOPS was equivalent to the direct connection structure. The two 0.4$\pi$ phase shifts applied to the 1st and 3rd TOPS act on the same path, results in a 0.8$\pi$ variation of the relative phase. To determine the changes in intensities due to 0$\pi$ or 0.8$\pi$, a 0.4$\pi$ phase-shift voltage was applied to 2$n$th TOPS, which changed the relative phase to $\theta _{2n}^{'} = {\theta _{2n}} + 0.4\pi $. Considering that the Jones vector after the (2$n$-1)th TOPS was $e_{(2 n-1) o}=\left[\left|e_{(2 n-1) 3}\right| e^{i \beta},\left|e_{(2 n-1) 4}\right|\right]=\left[e^{i \beta} /\sqrt{2}, 1 / \sqrt{2}\right]$, the intensity of the laser beam output from port 4 was
\begin{equation}
	\begin{gathered}
		{I_4} \hfill \\
		= \frac{1}{2}\left[ {1 - \left( {{{\left| {{e_{\left( {2n - 1} \right)4}}} \right|}^2} - {{\left| {{e_{\left( {2n - 1} \right)3}}} \right|}^2}} \right)\cos \left( {\theta _{2n}^{'}} \right)} \right. \hfill \\
		\left. { - 2\left| {{e_{\left( {2n - 1} \right)3}}} \right| \cdot \left| {{e_{\left( {2n - 1} \right)4}}} \right|\cos \left( \beta  \right)\sin \left( {\theta _{2n}^{'}} \right)} \right] \hfill \\
		= \frac{1}{2}\left[ {1 - \cos \left( \beta  \right)\sin \left( {\theta _{2n}^{'}} \right)} \right], \hfill \\ 
	\end{gathered} 
\end{equation}
where $\beta $ is the relative phase of the laser beams, and $\beta  \in \left[ {0,{\text{ 2}}\pi } \right)$.

When ${\theta _2} = 0$, $\beta  = {\pi  \mathord{\left/{\vphantom {\pi  {2 + m\pi }}} \right.\kern-\nulldelimiterspace} {2 + m\pi }}$, we have
\begin{equation}
	{I_4} = {1 \mathord{\left/{\vphantom {1 2}} \right.\kern-\nulldelimiterspace} 2}
\end{equation}

\indent When ${\theta _2} = \pi$, $\beta  = {\pi  \mathord{\left/{\vphantom {\pi  {2 + 0.8\pi  + m\pi }}} \right.\kern-\nulldelimiterspace} {2 + 0.8\pi  + m\pi }}$, we have
\begin{equation}
I_{4}=\frac{1}{2}[1-\sin (m \pi+0.8 \pi) \sin (0.4 \pi)]=0.7795 \text { or } 0.2205,
\end{equation}
where $m$ denotes the total number of phase shifters with a phase shift of $\pi $ from the 1st to $\left( {2n - 1} \right)$th TOPS. According to the value of ${I_4}$, we can determine the exact value of ${\theta _2}$.

\indent Using the aforementioned method, we calibrated the relative phase of the even-number TOPSs (2, 4, ..., $2n - 4$, $2n - 2$) from left to right when the laser beams transmitted from ports 2 to 4 (Fig. \ref{A1} (c)). Similarly, the relative phase of the odd-number TOPSs ($2n - 1$, $2n - 3$, ..., 5, 3) can be calibrated from right to left when the laser beams transmitted from ports 4 to 2. 
\begin{table}[tbph]
	\begin{center}
		\caption{Calibration results of 6-CPSs without constraint.} \label{tabA1}
		\renewcommand{\arraystretch}{1.3}
		\tabcolsep 12pt 
		\begin{tabular}{| c | c | c | c | c |}
			\hline
			\multirow{2}{*}{Result} & 2nd & 3rd & 4th & 5th \\
			{}  & TOPS & TOPS & TOPS & TOPS \\\hline
			$\theta $ (rad) & 0 & $\pi$ & $\pi$ & $\pi$  \\\hline
			${\theta _{th}}$ (rad) & 0.2177 & 3.0356 & 2.4491 & 1.9654  \\\hline
			$\Delta \theta $ (rad) & -0.2177 & 0.106 & 0.6925 & 1.1762 \\
			\hline 
		\end{tabular}
	\end{center}
\end{table} 

\indent We calibrated the 6-CPSs using the aforementioned method without constraint at 20 $^{\circ}$C. Table. \ref{tabA1} lists the calibration results of $\theta $, ${\theta _{th}}$ and $\Delta \theta $ from the 2nd to 5th TOPSs. The calibration results of $\Delta \theta $ and that in Table. \ref{tab2} were nearly the same.

\section{Method of calibrating ${\theta _j}$ in odd CPSs without constraint} \label{secB}
\setcounter{equation}{0}
\setcounter{figure}{0}
\setcounter{table}{0}
\renewcommand{\theequation}{B\arabic{equation}}
\renewcommand{\thefigure}{B\arabic{figure}}
\renewcommand{\thetable}{B\arabic{table}}

In the odd CPSs, the calibration method introduced below did not require a constraint when calibrating $\theta $ from the 2nd to $\left( {2n - 2} \right)$th TOPSs.

\indent The relative phase, $\theta $, of each TOPS can be calibrated to either 0 or $\pi$ using the odd CPSs calibration method without constraint. When all the relative phases are set to 0 or $\pi$, the MZI structures with the even TOPS were equivalent to the cross or direct-connection structures. The odd CPSs were equivalent to an MZI structure (Fig. \ref{B1} (a)). This was the initial state of our the following calibration.

\indent Similar to the method of calibrating the $\theta $ of even CPSs without constraint, we sequentially calibrate the $\theta $ of any even-number TOPSs (2, 4, ..., $2n - 4$, $2n - 2$) by applying 0.4$\pi$ phase shifts to the two adjacent odd TOPSs. As shown in Fig. \ref{B1} (b), to determine whether the relative phase, ${\theta _2}$, was 0 or $\pi$, we applied 0.4$\pi$ phase shifts to the 1st and 3rd TOPS, respectively. The Jones vector of the input beam from port 2 was ${e_{in}} = {\left[ {0,1} \right]^{\mathrm{T}}}$. The Jones vector after the $\left( {2n - 1} \right)$th TOPS was
\begin{equation}
e_{(2 n-1) o}=\left[\left|e_{(2 n-1) 3}\right| e^{i \beta},\left|e_{(2 n-1) 4}\right|\right]=\left[e^{i \beta} /\sqrt{2}, 1 / \sqrt{2}\right]
\end{equation} 

\indent The intensities of the laser beam output from ports 3 and 4 were
\begin{equation}
	\begin{gathered}
		{I_3} = \frac{1}{2}\left[ {1 + 2\left| {{e_{\left( {2n - 1} \right)3}}} \right|\left| {{e_{\left( {2n - 1} \right)4}}} \right|\sin \left( \beta  \right)} \right] = \frac{1}{2}\left[ {1 + \sin \left( \beta  \right)} \right], \hfill \\
		{I_4} = \frac{1}{2}\left[ {1 - 2\left| {{e_{\left( {2n - 1} \right)3}}} \right|\left| {{e_{\left( {2n - 1} \right)4}}} \right|\sin \left( \beta  \right)} \right] = \frac{1}{2}\left[ {1 - \sin \left( \beta  \right)} \right]. \hfill \\ 
	\end{gathered} 
\end{equation} 
\indent When ${\theta _2} = 0$, $\beta  = {\pi  \mathord{\left/{\vphantom {\pi  {2 + m\pi }}} \right.\kern-\nulldelimiterspace} {2 + m\pi }}$, we have
\begin{equation}
	{I_4} = {1 \mathord{\left/{\vphantom {1 2}} \right.\kern-\nulldelimiterspace} 2}
\end{equation}

\indent When ${\theta _2} = \pi$, $\beta  = {\pi  \mathord{\left/{\vphantom {\pi  {2 + 0.8\pi  + m\pi }}} \right.\kern-\nulldelimiterspace} {2 + 0.8\pi  + m\pi }}$, we have
\begin{equation}
	{I_4} = \frac{1}{2}\left[ {1 - \sin \left( {m\pi  + 0.8\pi } \right)} \right]{\text{ = 0}}{\text{.9045 or 0}}{\text{.0955}}{\text{.}}
\end{equation}
Here $m$ denotes the total number of phase shifters with a phase shift of $\pi$ from the 1st to $\left( {2n - 1} \right)$th TOPSs. According to the value of ${I_4}$, we calibrated the exact value of ${\theta _2}$. Using the aforementioned method, we calibrated the relative phase of the even-number TOPSs (2, 4, ..., $2n - 4$, $2n - 2$) from left to right when the laser beams transmitted from ports 2 to 4 (Fig. \ref{B1} (c)). Considering that the number of CPSs was odd, when the direction of light transmitted changed, we only calibrated $\theta $ of the even TOPSs. The odd-number TOPSs (3, 5, ..., $2n - 3$) could not be calibrated in this manner. 

\indent To calibrate the odd-number TOPSs, we added a ${\pi  \mathord{\left/{\vphantom {\pi  2}} \right.\kern-\nulldelimiterspace} 2}$ phase shift to the 1th TOPS. Thus, the first MZI structure was transformed into a $2 \times 2$ MMI with two phase delays, $\varphi $ (${\varphi  = {\pi  \mathord{\left/{\vphantom {\pi  2}} \right.\kern-\nulldelimiterspace} 2}}$ or ${\varphi  = {3\pi  \mathord{\left/{\vphantom {\pi  2}} \right.\kern-\nulldelimiterspace} 2}}$), placed before and after it, respectively (Fig. \ref{B1} (d)). After the transformation, the structure was similar to the even CPSs (Fig. \ref{A1} (b)) except for an additional phase shifter, $\varphi $, before the 2nd TOPS. When the input beam was ${e_{in}} = {\left[ {0{\text{ 1}}} \right]^{\mathrm{T}}}$ at port 2, after the equivalent $2 \times 2$ MMI and $\varphi $, the output beam became ${e_{2i}} = {\left[ {{e^{i\left( {{\pi  \mathord{\left/ {\vphantom {\pi  {2 + \varphi }}} \right.\kern-\nulldelimiterspace} {2 + \varphi }}} \right)}}/\sqrt 2 ,1/\sqrt 2 } \right]^{\mathrm{T}}}$. In this case, to calibrate the 3th TOPS, we applied 0.4$\pi$ phase shifts to the 2nd, 4th and $\left( {2n - 1} \right)$th TOPS. The Jones vector after the $\left( {2n - 2} \right)$th TOPS was denoted as   
\begin{equation}
	{e_{\left( {2n - 2} \right)o}} = \left[ {\left| {{e_{\left( {2n - 2} \right)3}}} \right|{e^{i\beta }},{\text{ }}\left| {{e_{\left( {2n - 2} \right)4}}} \right|} \right] = \left[ {{e^{i\beta }}/\sqrt 2 ,1/\sqrt 2 } \right].
\end{equation}

\indent The intensity of the laser beam output from port 4 was
\begin{equation}
	\begin{gathered}
		{I_4} \hfill \\
		= \frac{1}{2}\left[ {1 - \left( {{{\left| {{e_{\left( {2n - 2} \right)4}}} \right|}^2} - {{\left| {{e_{\left( {2n - 2} \right)3}}} \right|}^2}} \right)\cos \left( {\theta _{2n-1}^{'}} \right)} \right. \hfill \\
		\left. { - 2\left| {{e_{\left( {2n - 2} \right)3}}} \right| \cdot \left| {{e_{\left( {2n - 2} \right)4}}} \right|\cos \left( \beta  \right)\sin \left( {\theta _{2n-2}^{'}} \right)} \right] \hfill \\
		= \frac{1}{2}\left[ {1 + \cos \left( \beta  \right)\sin \left( {\theta _{2n-1}^{'}} \right)} \right], \hfill \\ 
	\end{gathered} 
\end{equation}
\indent When ${\theta _3} = 0$, $\beta  = \left( {l + 1} \right)\pi $, we have 
\begin{equation}
	{I_4} = {\text{0}}{\text{.9755 or }}0.0245.
\end{equation}

\indent When ${\theta _3} = \pi$, $\beta  = \left( {l + 1} \right)\pi  + 0.8\pi $, we have
\begin{equation}
	{\text{ }}{I_4} = {\text{0}}{\text{.8847 or }}0.1153.
\end{equation}
Here $l$ represents the total number of phase shifters with a value of $\pi$ from the 2nd to $\left( {2n - 2} \right)$th TOPS. Subsequently, the odd-number TOPSs (3, 5, ..., $2n - 3$) were calibrated (Fig. \ref{B1} (e)). Finally, we achieve the calibration of  $\theta $ from the 2nd to $\left( {2n - 2} \right)$th TOPSs in odd CPSs without constraint.

\section{The fidelity considering the biased splitting ratios ans unbalanced losses} \label{secC}
\setcounter{equation}{0}
\setcounter{figure}{0}
\setcounter{table}{0}
\renewcommand{\theequation}{C\arabic{equation}}
\renewcommand{\thefigure}{C\arabic{figure}}
\renewcommand{\thetable}{C\arabic{table}}

To evaluate the effect of biased splitting ratios and unbalanced losses, we construct a MMI with a splitting ratio $\eta $, and attenuation constants $\tau $ and $\kappa $, respectively. In this case, the matrix of MMI can be expressed as
\begin{equation}
\begin{aligned}
	M_{\mathrm{MMI}}(\eta) & =\left[\begin{array}{cc}
		e^{-\tau / 2} & 0 \\
		0 & e^{-\kappa / 2}
	\end{array}\right] \cdot\left[\begin{array}{cc}
		\sqrt{\eta} & i \sqrt{1-\eta} \\
		i \sqrt{1-\eta} & \sqrt{\eta}
	\end{array}\right] \\
	& =\left[\begin{array}{cc}
		e^{-\tau / 2} \sqrt{\eta} & i e^{-\tau / 2} \sqrt{1-\eta} \\
		i e^{-\kappa / 2} \sqrt{1-\eta} & e^{-\kappa / 2} \sqrt{\eta}
	\end{array}\right] .
\end{aligned}
\end{equation}
When the Jones vector of the input beam is ${e_{in}} = {\left[ {0,1} \right]^{\mathrm{T}}}$, the Jones vector of the output beam is
\begin{equation}
{e_{out}} = {M_{{\text{MMI}}}}\left( \eta  \right) \cdot {e_{in}} = \left[ {\begin{array}{*{20}{c}}
		{i{e^{{{ - \tau } \mathord{\left/
							{\vphantom {{ - \tau } 2}} \right.
							\kern-\nulldelimiterspace} 2}}}\sqrt {1 - \eta } } \\ 
		{{e^{{{ - \kappa } \mathord{\left/
							{\vphantom {{ - \kappa } 2}} \right.
							\kern-\nulldelimiterspace} 2}}}\sqrt \eta  } 
\end{array}} \right].
\end{equation} 
The intensities of laser beams output from ports 3 and 4 are
 \begin{equation}
\begin{gathered}
	{T_{32}} = {e^{ - \tau }} \cdot \left( {1 - \eta } \right), \hfill \\
	{T_{42}} = {e^{ - \kappa }} \cdot \eta . \hfill \\ 
\end{gathered} 
 \end{equation}
The imbalance degree, ${\Delta _2}$, of the MMI can be defined as 
 \begin{equation}
{\Delta _2} = \left| {10 \cdot {{\log }_{10}}\left( {{{{T_{32}}} \mathord{\left/
				{\vphantom {{{T_{32}}} {{T_{42}}}}} \right.
				\kern-\nulldelimiterspace} {{T_{42}}}}} \right)} \right| = \left| {10 \cdot {{\log }_{10}}\left[ {\frac{{{r^2}\left( {1 - \eta } \right)}}{\eta }} \right]} \right|,
\end{equation}
where $r = {e^{{{ - \left( {\tau  - \kappa } \right)} \mathord{\left/
{\vphantom {{ - \left( {\alpha  - \kappa } \right)} 2}} \right.
\kern-\nulldelimiterspace} 2}}}$. For MZI or 1-CPS structure constructed by above MMI, when the Jones vector of input beam is ${e_{in}} = {\left[ {0,1} \right]^{\mathrm{T}}}$, the Jones vector of output beam is 
 \begin{equation}
\begin{aligned}
	e_{\text {out }} & =\left[\begin{array}{l}
	e_{3} \\
	e_{4}
\end{array}\right]=M_{\mathrm{MZI}}(\eta) \cdot e_{\text {in }} \\
& =\left[\begin{array}{c}
	i e^{-\tau / 2}\left(e^{-\kappa / 2}+e^{-\tau / 2} e^{i \theta}\right) \sqrt{(1-\eta) \eta} \\
	e^{-\kappa / 2}\left(e^{-\tau / 2} e^{i \theta}(\eta-1)+e^{-\kappa / 2} \eta\right)
\end{array}\right].
\end{aligned}
\end{equation}
Their corresponding intensities are 
\begin{equation}
{I_3} = {e^{ - \tau }}\left( {1 - \eta } \right)\eta \left[ {{e^{ - \tau }} + {e^{ - \kappa }} + 2{e^{ - {{\left( {\tau  + \kappa } \right)} \mathord{\left/
					{\vphantom {{\left( {\tau  + \kappa } \right)} 2}} \right.
					\kern-\nulldelimiterspace} 2}}}\cos \left( \theta  \right)} \right],{\rm{    }}
\end{equation}
\begin{equation}
{I_4} = {e^{ - \kappa }}\left[ {{e^{ - \tau }}{{\left( {1 - \eta } \right)}^2} + {e^{ - \kappa }}{\eta ^2} - 2{e^{ - {{\left( {\tau  + \kappa } \right)} \mathord{\left/
					{\vphantom {{\left( {\tau  + \kappa } \right)} 2}} \right.
					\kern-\nulldelimiterspace} 2}}}\left( {1 - \eta } \right)\eta \cos \left( \theta  \right)} \right].{\rm{   }}
\end{equation}
The extinction ratio of the ports 3 and 4 can be expressed as
\begin{equation}
\begin{aligned}
	E R_{\text {port3 }} & =10 \cdot \log _{10}\left(i_{3 \max } / i_{3 \min }\right) \\
& =10 \cdot \log _{10}\left(\frac{e^{-\tau / 2}+e^{-\kappa / 2}}{e^{-\tau / 2}-e^{-\kappa / 2}}\right)^{2}=10 \cdot \log _{10}\left(\frac{r+1}{r-1}\right)^{2},
\end{aligned}
\end{equation}
\begin{equation}
\begin{gathered}
	E{R_{{\text{port4}}}} = 10 \cdot {\log _{10}}\left( {{{{i_{4\max }}} \mathord{\left/
				{\vphantom {{{i_{4\max }}} {{i_{4\min }}}}} \right.
				\kern-\nulldelimiterspace} {{i_{4\min }}}}} \right) \hfill \\
	= 10 \cdot {\log _{10}}{\left[ {\frac{{{e^{{{ - \tau } \mathord{\left/
									{\vphantom {{ - \tau } 2}} \right.
									\kern-\nulldelimiterspace} 2}}}\left( {1 - \eta } \right) + {e^{{{ - \kappa } \mathord{\left/
									{\vphantom {{ - \kappa } 2}} \right.
									\kern-\nulldelimiterspace} 2}}}\eta }}{{{e^{{{ - \tau } \mathord{\left/
									{\vphantom {{ - \tau } 2}} \right.
									\kern-\nulldelimiterspace} 2}}}\left( {1 - \eta } \right) - {e^{{{ - \kappa } \mathord{\left/
									{\vphantom {{ - \kappa } 2}} \right.
									\kern-\nulldelimiterspace} 2}}}\eta }}} \right]^2} = 10 \cdot {\log _{10}}{\left[ {\frac{{r\left( {1 - \eta } \right) + \eta }}{{r\left( {1 - \eta } \right) - \eta }}} \right]^2}. \hfill \\ 
\end{gathered} 
\end{equation}
\begin{figure}[htbp]
	\vspace{-0.3cm}
	\centering
	\includegraphics[scale=0.55]{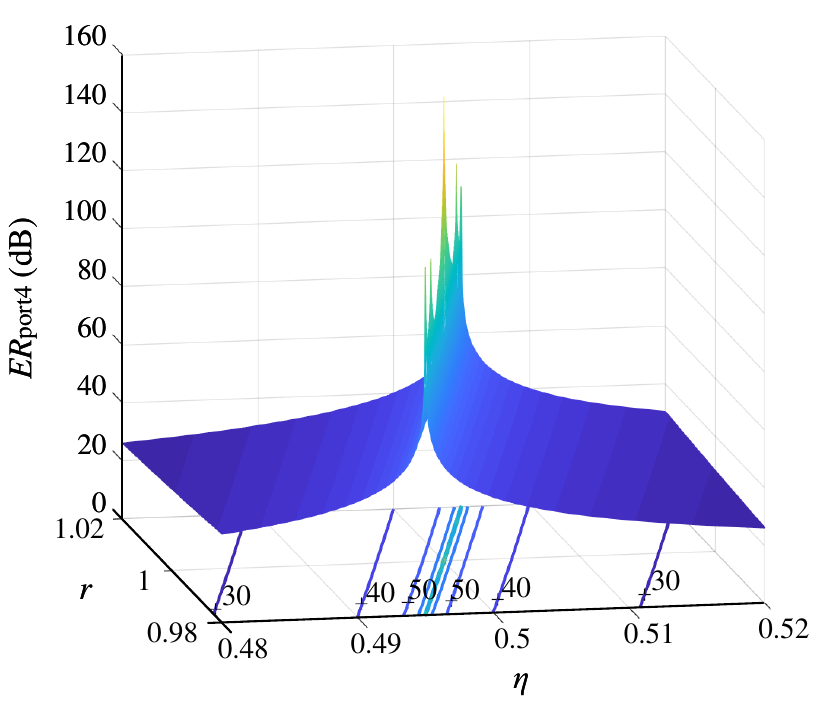}
	\caption{The contours at different extinction ratios of $E{R_{{\text{port}}4}}$.} 
	\label{figC1}
\end{figure}  

\indent From the expression of $E{R_{{\text{port3}}}}$, we can see that it has no relationship with $\eta $. When the ratio $r $ approaches to 1, the $E{R_{{\text{port3}}}}$ approached to higher and infinite. The dependence of $E{R_{{\text{port4}}}}$ on $r $ and $\eta $ is shown in Fig. \ref{figC1}, the contours of different extinction ratios are plotted. For a constant value of $E{R_{{\text{port4}}}}$, $r $ and $\eta $ has two nearly linear relationships.
\begin{figure}[htbp]
	\centering
	\includegraphics[scale=0.55]{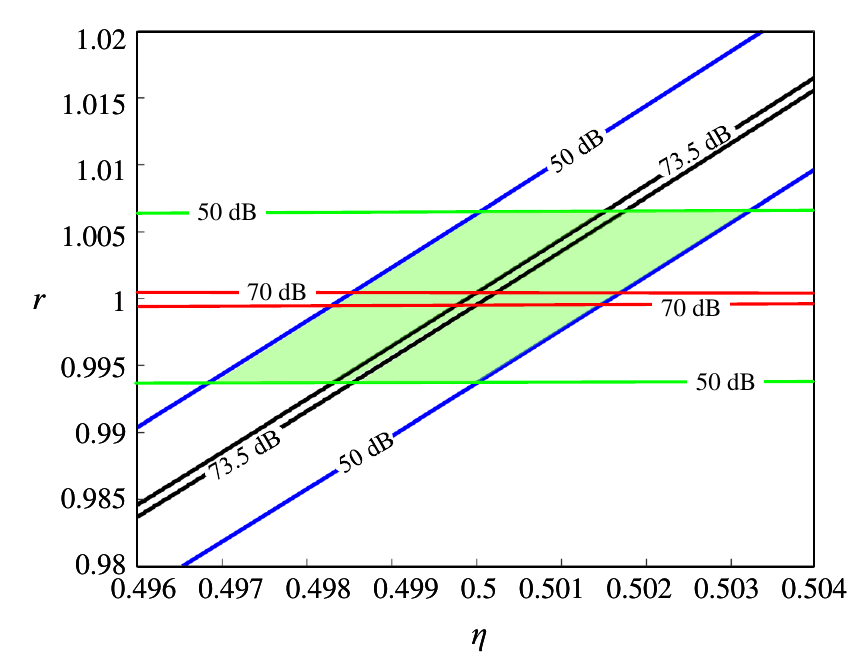}
	\caption{The constrained transmission efficiency $\eta $ and ratio $r $.} 
	\label{figC2}
\end{figure} 

\indent From the simulation results, we can obtain the transmission efficiencies ${T_{32}}$ and ${T_{42}}$. In this case, the extinction ratio $E{R_{{\text{port4}}}} = 73.5{\text{ dB}}$, the relationship of $\eta $ and $r $ is represented by the black solid lines in Fig. \ref{figC2}. The relationship of $\eta $ and $r $ at the $E{R_{{\text{port4}}}} = 50{\text{ dB}}$ is represented by blue lines. The $E{R_{{\text{port3}}}}$ is only related to $r $. The values of $r $ at $E{R_{{\text{port3}}}} = 50{\text{ dB}}$, $70{\text{ dB}}$ are represented by the green lines and red lines, respectively. Obviously, we can use the extinction ratio to restrict the range of $\eta $ and $r $. For example, $E{R_{{\text{port3}}}} = E{R_{{\text{port4}}}} \geqslant 50{\text{ dB}}$ will constraint $\eta $ and $r $ in the green area. In this case, the minimum fidelity is 99.991{$\%$}. When the extinction ratio increases, the ratio $r $, the splitting ratio $\eta $ and the fidelity approaches to the ideal value of 1, 0.5, and 1, respectively.

\indent From the above analysis, we can see that a measured extinction ratio greater than 50 dB can ensure the high fidelity greater than 99.991{$\%$}. The mature fabrication technology can make sure the consistency of the MZI structures in the 6-CPSs. Therefore, a fidelity greater than 99.97{$\%$} is achieved in our experiment.


\end{document}